\documentclass[longauth]{aa}

\usepackage{graphicx}
\usepackage{txfonts}
\usepackage[colorlinks=true,citecolor=blue,linkcolor=black]{hyperref}
\usepackage{siunitx}
\usepackage{xcolor}
\usepackage{float}
\usepackage{bm}
\usepackage{soul,xcolor}
\setstcolor{red}
\definecolor{alizarin}{rgb}{0.82, 0.1, 0.26}

\newcommand{\oiii}{\textup{[O\,\textsc{iii}]}}
\newcommand{\nii}{\textup{[N\,\textsc{ii}]}}
\newcommand{\sii}{\textup{[S\,\textsc{ii}]}}
\newcommand{\oi}{\textup{[O\,\textsc{i}]}}
\newcommand{\siii}{\textup{[S\,\textsc{iii}]}}
\newcommand{\oii}{\textup{[O\,\textsc{ii}]}}
\newcommand{\hii}{\textup{H}\,\textsc{ii}}
\newcommand{\ha}{\textup{H}\ensuremath{\alpha}}
\newcommand{\hb}{\textup{H}\ensuremath{\beta}}

\newcommand{\SHa}{\ensuremath{\Sigma_{\mathrm{H\alpha}}}}

\begin{document}

   \title{A tale of two DIGs: \\The relative role of \hii\ regions and low-mass hot evolved stars in powering the diffuse ionised gas (DIG) in PHANGS--MUSE galaxies}
     
\author{Francesco~Belfiore\inst{\ref{arcetri}}
        \and
Francesco~Santoro\inst{\ref{mpia}}
\and
Brent~Groves\inst{\ref{icrar},\ref{anu}}
\and
Eva~Schinnerer\inst{\ref{mpia}}
\and
Kathryn~Kreckel\inst{\ref{rechen}}
\and
Simon~C.~O.~Glover\inst{\ref{zah}}
\and
Ralf~S.~Klessen\inst{\ref{zah},\ref{zw}}
 \and
Eric Emsellem\inst{\ref{eso},\ref{lyon}}
\and
Guillermo~A.~Blanc\inst{\ref{uch},\ref{carn}}
\and 
Enrico Congiu\inst{\ref{uch}} 
\and
Ashley~T.~Barnes\inst{\ref{UBonn}} 
\and
Médéric~Boquien\inst{\ref{UA}}
\and
M\'elanie~Chevance\inst{\ref{rechen}}
\and
Daniel~A.~Dale\inst{\ref{wyo}} 
\and
J.~M.~Diederik~Kruijssen\inst{\ref{rechen}} 
\and
Adam~K.~Leroy\inst{\ref{ohio}}
\and
Hsi-An~Pan\inst{\ref{mpia},\ref{taiwan}}
\and
Ismael~Pessa\inst{\ref{mpia}}
\and
Andreas~Schruba \inst{\ref{mpe}}
\and
Thomas~G.~Williams\inst{\ref{mpia}}
}

\institute{INAF — Osservatorio Astrofisico di Arcetri, Largo E. Fermi 5, I-50125, Florence, Italy\label{arcetri}\\
   \email{francesco.belfiore@inaf.it}
    \and Max-Planck-Institute for Astronomy, K\"onigstuhl 17, D-69117 Heidelberg, Germany\label{mpia}
    \and International Centre for Radio Astronomy Research, University of Western Australia, 35 Stirling Highway, Crawley, WA 6009, Australia\label{icrar}
    \and Research School of Astronomy and Astrophysics, Australian National University, Canberra, ACT 2611, Australia\label{anu}
    \and Astronomisches Rechen-Institut, Zentrum f\"ur Astronomie der Universit\"at Heidelberg, M\"onchhofstra{\ss}e 12-14, D-69120 Heidelberg, Germany\label{rechen}
    \and Universit\"at Heidelberg, Zentrum f\"ur Astronomie, Institut f\"ur theoretische Astrophysik, Albert-Ueberle-Stra{\ss}e 2, D-69120, Heidelberg, Germany\label{zah}
    \and Universit\"at Heidelberg, Interdisziplin\"ares Zentrum f\"ur Wissenschaftliches Rechnen, Im Neuenheimer Feld 205, D-69120 Heidelberg, Germany\label{zw}
    \and European Southern Observatory, Karl-Schwarzschild-Stra{\ss}e 2, 85748 Garching, Germany\label{eso}
    \and Univ Lyon, Univ Lyon1, ENS de Lyon, CNRS, Centre de Recherche Astrophysique de Lyon UMR5574, F-69230 Saint-Genis-Laval France\label{lyon}
    \and Departamento de Astronom\'ia, Universidad de Chile, Santiago, Chile\label{uch}
    \and Observatories of the Carnegie Institution for Science, Pasadena, CA, USA\label{carn}
    \and Argelander-Institut f\"ur Astronomie, Universit\"at Bonn, Auf dem H\"ugel 71, D-53121 Bonn, Germany\label{UBonn}
    \and Centro de Astronomía (CITEVA), Universidad de Antofagasta, Avenida Angamos 601, Antofagasta, Chile\label{UA}
    \and Department of Physics and Astronomy, University of Wyoming, Laramie, WY 82071, USA\label{wyo}
    \and Department of Astronomy, The Ohio State University, 140 West 18th Avenue, Columbus, OH 43210, USA\label{ohio}
    \and Department of Physics, Tamkang University, No.151, Yingzhuan Rd., Tamsui Dist., New Taipei City 251301, Taiwan \label{taiwan}
    \and Max-Planck-Institute for extraterrestrial Physics, Giessenbachstra{\ss}e 1, D-85748 Garching, Germany\label{mpe}
}

\date{Received XXX; accepted XXX}

 
\abstract{
We use integral field spectroscopy from the PHANGS--MUSE survey, which resolves the ionised interstellar medium structure at ${\sim}50$~pc resolution in 19 nearby spiral galaxies, to study the origin of the diffuse ionised gas (DIG). We examine the physical conditions of the diffuse gas by first removing morphologically defined \hii\ regions and then binning the low-surface-brightness areas to achieve significant detections of the key nebular lines in the DIG. 
A simple model for the leakage and propagation of ionising radiation from \hii\ regions is able to reproduce the observed distribution of \ha\ in the DIG. This model infers a typical mean free path for the ionising radiation of 1.9~kpc for photons propagating within the disc plane. Leaking radiation from \hii\ regions also explains the observed decrease in line ratios of low-ionisation species (\sii/\ha, \nii/\ha, and \oi/\ha) with increasing \ha\ surface brightness (\SHa). Emission from hot low-mass evolved stars, however, is required to explain: (1) the enhanced low-ionisation line ratios observed in the central regions of some of the galaxies in our sample; (2) the observed trends of a flat or decreasing \oiii/\hb\ with \SHa; and (3) the offset of some DIG regions from the typical locus of \hii\ regions in the Baldwin--Phillips--Terlevich (BPT) diagram, extending into the area of low-ionisation (nuclear) emission-line regions (LI[N]ERs). Hot low-mass evolved stars make a small contribution to the energy budget of the DIG (2\% of the galaxy-integrated \ha\ emission), but their harder spectra make them fundamental contributors to \oiii\ emission. The DIG might result from a superposition of two components, an energetically dominant contribution from young stars and a more diffuse background of harder ionising photons from old stars. This unified framework bridges observations of the Milky Way DIG with LI(N)ER-like emission observed in nearby galaxy bulges.}
\titlerunning{A tale of two DIGs}
\authorrunning{F. Belfiore}
\maketitle


\section{Introduction}
\label{sec:intro}

\hii\ regions are key tracers of the rate of star formation and the degree of chemical enrichment of the interstellar medium (ISM) in galaxies across the entire range of cosmological distances accessible to current instrumentation \citep{Maiolino2019, Kewley2019}. Detailed studies of \hii\ regions in nearby galaxies have quantified the effect of feedback from massive stars on their environment (e.g. \citealt{Lopez2011,Pellegrini2012a,Lopez2014, McLeod2019a,McLeod2020, Barnes2020, Olivier2021, DellaBruna2021}) and allowed the development of empirical relations that describe the conversion of cold gas into stars (e.g. \citealt{Kennicutt1989b,Schinnerer2019, Chevance2020}). Feedback from massive stars, in the form of radiation, winds, and supernovae, is a crucial ingredient in determining the state of the ISM in which new stars form \citep{Hopkins2014, Klessen2016, Gatto2017}.

A large fraction of the ionising radiation from massive stars may escape from \hii\ regions in their immediate vicinity and contribute to powering the diffuse ionised gas (DIG): a tenuous plasma with low electron densities ($\rm n_{e} \sim 10^{-1} {-} 10^{-2}~cm^{-3}$) and temperatures slightly higher than those of typical \hii\ regions (${\sim}10^{4}$\,K). Understanding what fraction of the DIG is powered by leaking radiation from \hii\ regions, and therefore the ionising photon escape fraction from regions of star formation, is of fundamental importance for drawing a full budget of the feedback mechanisms acting to disrupt giant molecular clouds, and therefore driving the timescales of the cycle of star formation in galaxies \citep{Chevance2022}.
The distribution of ionised hydrogen in the DIG is traced by the \ha\ recombination line. In general, the DIG is found to contain roughly half of the total \ha\ emission in galaxies \citep{Oey2007, Kreckel2016, Chevance2020}.
This emission is often observed to be spatially associated with nearby \hii\ regions \citep{Ferguson1996, Zurita2000, Zurita2002, Madsen2006, Seon2009, Howard2018, Pellegrini2020}, supporting the idea that the DIG may be ionised by leaking radiation. Detailed assessments of the energetics involved support this argument for the DIG in the Milky Way \citep{Reynolds1990, McKee1997, Haffner2009}.

In external galaxies, collisionally excited optical line ratios constitute our main window into the physical conditions of the DIG and can shed light onto its ionisation source. In nearby edge-on galaxies \citep{Otte2001, Jones2017, Levy2019}, spectroscopic observations show an increase in emission of low-ionisation species in the DIG with respect to \hii\ regions. In particular, the line ratios \sii$\lambda\lambda$6717,31/\ha, \nii$\lambda$6584/\ha, and \oi$\lambda$6300/\ha\ are found to increase with scale height from the midplane, or as a function of decreasing \ha\ surface brightness (\SHa; \citealt{Hill2014}).
Models for gas photoionised by leaking radiation from \hii\ regions in the low-density regime are able to explain the increase in low-ionisation line ratios such as \sii/\ha, \nii/\ha, and \oi/\ha.
Furthermore, intervening absorption from neutral gas preferentially absorbs photons near the hydrogen ionisation potential, therefore hardening the ionising spectrum, which can also lead to an increase in low-ionisation line ratios \citep{Wood2004}.

However, leaking \hii\ region models are generally unable to reproduce the relative strengths of high-ionisation lines (e.g. \oiii$\lambda$5007/\hb) in the extraplanar DIG without additional sources of heating \citep{Rand1998, Reynolds1999}. Several physical processes have been suggested as sources for an additional heating term (e.g. shocks: \citealt{Collins2001}, turbulence:  \citealt{Binette2009}, or magnetic reconnection: \citealt{Lazarian2020}). An alternative solution, provided by \cite{Flores-Fajardo2011} in a study of the prototypical edge-on galaxy NGC~891, is that the DIG emission can be modelled as a combination of photoionisation from leaky \hii\ regions and hot low-mass evolved stars (HOLMES). HOLMES are low- to intermediate-mass stars ($0.8{-}8.0~M_\odot$) in all the stages of stellar evolution subsequent to the asymptotic giant branch (post-AGB), including the hydrogen-burning, white-dwarf-cooling, and intermediate phases \citep{Stanghellini2000, Stasinska2008}. The term `post-AGB' has been used in other studies \citep{Binette1994, Belfiore2016a, Byler2019} to refer to the same population of sources. HOLMES have a harder ionising spectrum than \hii\ regions and are therefore capable of producing higher \oiii/\hb\ ratios, even with relatively weak radiation fields. HOLMES are a prime candidate to explain the \oiii/\hb\ line ratios in the DIG of star-forming galaxies because, unlike the other potential candidates, their presence in galaxies is undisputed.

Data from kiloparsec-resolution integral field spectroscopy (IFS) surveys of nearby galaxies, such as CALIFA (Calar Alto Legacy Integral Field spectroscopy Area survey, \citealt{Sanchez2012a}), SAMI (Sydney-Australian-Astronomical-Observatory Multi-object Integral-Field Spectrograph, \citealt{Croom2012}), and MaNGA (Mapping Nearby Galaxies at APO, \citealt{Bundy2015}), has demonstrated the importance of photoionisation by HOLMES in early-type galaxies \citep{Sarzi2010, Singh2013a} and central regions of massive late-type galaxies \citep{Belfiore2016a}, which are devoid of recent star formation. These systems show line ratios that fall in the low-ionisation (nuclear) emission-line region (LI[N]ER) corner of classical Baldwin--Phillips--Terlevich (BPT;  \citealt{Baldwin1981, Kewley2001, Kewley2006}) diagrams, as expected from ionisation sources other than young stars.

There is also increasing evidence with regard to the importance of HOLMES in explaining the observed line ratios in the DIG of star-forming galaxies. \cite{Zhang2017} invoke a contribution from HOLMES to explain the high \oiii/\hb\ line ratios observed at low \SHa\ in star-forming galaxies from the MaNGA survey. Other authors \citep{Lacerda2018, ValeAsari2019,Espinosa-Ponce2020} use data at kiloparsec resolution to effectively redefine the DIG as regions with a low equivalent width of \ha\ emission (e.g. ${\rm EW}(\text{\ha}) < 3$~\AA), which models suggest may be consistent with ionisation from HOLMES. However, this redefinition is problematic for studies interested in determining the ionisation sources of the DIG since it defines the DIG a priori as the ionised gas component consistent with ionisation from HOLMES. Unfortunately, at the typical resolution provided by CALIFA, SAMI, or MaNGA  (${\sim}0.8$ for CALIFA to $2$~kpc for MaNGA), regions classified as star-forming consist of an agglomeration of several bright \hii\ regions \citep{Mast2014, Poetrodjojo2019}, and it is not possible to spatially resolve these regions from any nearby, potentially associated, DIG.

This uncertainty in identifying the DIG leads to disagreement over the importance of the DIG in determining chemical abundances. Abundance measurements rely on photoionisation models, which do not generally include DIG contamination, or on empirical relations calibrated on bright \hii\ regions (themselves generally expected to suffer only negligible DIG contamination). The \nii/\ha\ line ratio, for example, is sometimes used as a metallicity tracer, especially at high redshift \citep{pettini2004}. 
Considering that roughly half of the \ha\ emission may originate outside \hii\ regions, and that the \nii/\ha\ ratio is enhanced in the DIG, the abundance derived without taking the DIG into account inevitably results in a biased measurement (e.g. \citealt{Zhang2017, Poetrodjojo2019}). 
A reassessment of the relative importance of leaking radiation and HOLMES in powering the DIG in star-forming disc galaxies is therefore urgently needed to make progress in quantifying the \hii\ regions' ionising photon escape fraction and to measure chemical abundances accurately.

Making such advances would require sensitive, high-spatial-resolution spectral mapping of nearby star-forming galaxies. The ideal dataset would have sufficient spatial resolution to resolve the average separation between \hii\ regions (${\sim}100 {-} 200$~pc; \citealt{Chevance2020}) in order to allow \hii\ region emission to be spatially separated from the DIG.

\begin{table*}
\caption{{Key properties of the PHANGS--MUSE sample of galaxies used in this work.} Distances are taken from the compilation of \cite{Anand2021}. Stellar masses, star-formation rate (SFR), and offsets from the star-formation main sequence ($\Delta_{\rm SFMS}$) are taken from the PHANGS sample paper \citep{Leroy2021a} and are based on a combined \textit{WISE}+\textit{GALEX} analysis \citep{Salim2018, Leroy2019}.
$R_{25}$ is the $B$-band isophotal radius recovered via HyperLEDA \citep{Makarov2014}. Position angle (PA) and inclination ($i$) are taken from analysis of the CO velocity fields from \cite{Lang2020}. The median (maximum and minimum) PSF across each MUSE mosaic is taken from \citealt{Emsellem2021}. The integrated fraction of extinction-corrected \ha\ emission in the DIG ($f_\mathrm{DIG}$) is computed in Sect.~\ref{sec:ha_in_DIG}.}
\label{tab:tab1}
\centering
\begin{tabular}{lrcccccccccc}
\hline \hline
Name & Distance & ${\rm Log}(M_\star) $ & $\rm Log(SFR) $ & $\rm \Delta_{SFMS} $ & $\rm R_{25} $ & PA & $i$ & pc/$''$ & Survey Areas & PSF & $f_\mathrm{DIG}$\\
     & Mpc & [$M_\odot$]        & [$M_\odot~\mathrm{yr}^{-1}$]  & dex   & arcmin  & deg  & deg & & kpc$^2$ & arcsec & \\
\hline
\hline
NGC~0628 & 9.8 & 10.34 & 0.24 & 0.18 & 4.9 & 20.7 & 8.9 & 47.7 & 89 & 0.73$^{+0.11}_{-0.13}$ & 0.45 \\
NGC~1087 & 15.9 & 9.93 & 0.12 & 0.33 & 1.5 & 359.1 & 42.9 & 76.8 & 126 & 0.74$^{+0.10}_{-0.12}$ & 0.38\\
NGC~1300 & 19.0 & 10.62 & 0.07 & $-$0.18 & 3.0 & 278.0 & 31.8 & 92.1 & 356 & 0.63$^{+0.18}_{-0.13}$& 0.52 \\
NGC~1365 & 19.6 & 10.99 & 1.23 & 0.72 & 6.0 & 201.1 & 55.4 & 94.9 & 409 & 0.82$^{+0.26}_{-0.24}$ & 0.19\\
NGC~1385 & 17.2 & 9.98 & 0.32 & 0.50 & 1.7 & 181.3 & 44.0 & 83.5 & 101 & 0.49$^{+0.10}_{-0.11}$ & 0.30\\
NGC~1433 & 18.6 & 10.87 & 0.05 & $-$0.36 & 3.1 & 199.7 & 28.6 & 90.3 & 426 & 0.65$^{+0.18}_{-0.14}$ & 0.53\\
NGC~1512 & 18.8 & 10.71 & 0.11 & $-$0.21 & 4.2 & 261.9 & 42.5 & 91.3 & 266 & 0.80$^{+0.38}_{-0.16}$ & 0.48\\
NGC~1566 & 17.7 & 10.78 & 0.66 & 0.29 & 3.6 & 214.7 & 29.5 & 85.8 & 208 & 0.64$^{+0.09}_{-0.10}$ & 0.33\\
NGC~1672 & 19.4 & 10.73 & 0.88 & 0.56 & 3.1 & 134.3 & 42.6 & 94.1 & 250 & 0.72$^{+0.17}_{-0.08}$ & 0.23\\
NGC~2835 & 12.2 & 10.00 & 0.09 & 0.26 & 3.2 & 1.0 & 41.3 & 59.2 & 87 & 0.85$^{+0.23}_{-0.18}$ & 0.43\\
NGC~3351 & 10.0 & 10.36 & 0.12 & 0.05 & 3.6 & 193.2 & 45.1 & 48.3 & 73 & 0.74$^{+0.24}_{-0.13}$ & 0.30\\
NGC~3627 & 11.3 & 10.83 & 0.58 & 0.19 & 5.1 & 173.1 & 57.3 & 54.9 & 85 & 0.77$^{+0.21}_{-0.10}$ & 0.32\\
NGC~4254 & 13.1 & 10.42 & 0.49 & 0.37 & 2.5 & 68.1 & 34.4 & 63.5 & 169 & 0.58$^{+0.23}_{-0.14}$ & 0.43\\
NGC~4303 & 17.0 & 10.52 & 0.73 & 0.54 & 3.4 & 312.4 & 23.5 & 82.4 & 214 & 0.58$^{+0.12}_{-0.07}$ & 0.36\\
NGC~4321 & 15.2 & 10.75 & 0.55 & 0.21 & 3.0 & 156.2 & 38.5 & 73.7 & 191 & 0.64$^{+0.45}_{-0.18}$ & 0.41\\
NGC~4535 & 15.8 & 10.53 & 0.33 & 0.14 & 4.1 & 179.7 & 44.7 & 76.5 & 124 & 0.44$^{+0.03}_{-0.01}$ & 0.37\\
NGC~5068 & 5.2 & 9.40 & -0.56 & 0.02 & 3.7 & 342.4 & 35.7 & 25.2 & 23 & 0.73$^{+0.23}_{-0.21}$ & 0.32\\
NGC~7496 & 18.7 & 10.00 & 0.35 & 0.53 & 1.7 & 193.7 & 35.9 & 90.8 & 92 & 0.79$^{+0.03}_{-0.17}$ & 0.37\\
IC~5332 & 9.0 & 9.67 & -0.39 & 0.01 & 3.0 & 74.4 & 26.9 & 43.7 & 34 & 0.72$^{+0.08}_{-0.12}$ & 0.55\\
\hline
\end{tabular}
\end{table*}

A dataset suitable for a high-spatial-resolution study of the DIG is now available thanks to the Physics at High Angular resolution in Nearby GalaxieS (PHANGS)\footnote{\url{www.phangs.org}} project, which includes mapping of 19 nearby galaxies at ${<}100$~pc resolution obtained via a Large Programme (PI: E.~Schinnerer) with the MUSE (Multi Unit Spectroscopic Explorer) integral field spectrograph on the ESO/Very Large Telescope (VLT) \citep{Emsellem2021}. 
In this paper we make use of the PHANGS--MUSE dataset to study the spatial structure of the DIG and its line ratios. 
In Sect.~\ref{sec:data_and_sample} we describe the data and our analysis tools and present our methodology for separating \hii\ regions from DIG based on \ha\ morphology. We present the general characteristics of the DIG in our galaxy sample in Sect.~\ref{sec:gen_cat}.
We then test the canonical assumption that the DIG is powered by leaked radiation from \hii\ regions by comparing our observations with simple predictions of the spatial distribution of the DIG under this scenario (Sect.~\ref{sec:DIG_spatial_model}). We describe the change in DIG line ratios as a function of proxies for metallicity and the ionisation parameter in Sect.~\ref{sec:line_ratios} and explain the trends via photoionisation models. In Sect.~\ref{sec:discussion} we present our main take-away points with regards to the powering source of the DIG and what role HOLMES play in its ionisation state.

For the rest of the paper we make use of the following shorthand notation:
$\nii \equiv \nii\lambda6584$;
$\sii \equiv \sii\lambda6717 + \sii\lambda6731$;
$\oiii \equiv \oiii\lambda5007$; and
$\oi \equiv \oi\lambda6300$.

\section{Sample and data analysis}
\label{sec:data_and_sample}

\subsection{Observations and data reduction}
\label{sec:ObsDRS}

PHANGS--MUSE consists of 19 galaxies, selected to be nearby ($D < 20$~Mpc), close to the star-formation main sequence, and moderately inclined. The main properties of the sample, which spans the stellar mass range $\log(M_\star/M_{\odot}) = 9.4 {-} 11.0$, are summarised in Table~\ref{tab:tab1}. A description of the PHANGS--MUSE sample selection, observations, and associated data is presented in \cite{Emsellem2021}. In the rest of this section we offer a brief summary.

The MUSE IFS instrument \citep{Bacon2010} at the ESO/VLT was used to cover the star-forming part of the disc. For each galaxy we obtain a contiguous mosaic of several $1' \times 1'$ MUSE pointings (between 3 and 15 pointings per galaxy). The majority of the data was obtained via a dedicated ESO Large Programme  (PI: E.~Schinnerer), running from ESO periods 100--106 (October 2017--March 2021). Three extra MUSE pointings (one each in NGC~1566, NGC~1385 and NGC~4321) are not included here because their reduction was pending during the development of this work, but they are included in the PHANGS public data release. Eight galaxies in the sample were observed by making use of the MUSE wide-field ground layer adaptive optics mode, while the remaining ones were observed in natural seeing.
All the data were reduced via the \textsc{pymusepipe} python package\footnote{Available from \url{https://github.com/emsellem/pymusepipe}}, which wraps the \textsc{esorex} MUSE reduction recipes \citep{Weilbacher2020a} and additionally provides routines for astrometric alignment, mosaicking and point spread function (PSF) homogenisation. 

Astrometry and flux calibration were performed by comparing the MUSE data with ESO/MPG 2.2m Wide Field Imager and LCO/DuPont Direct CCD (for NGC~7496 only) broadband images (Razza et al., in preparation), whose astrometry was matched to that of \textit{Gaia} data release~2 \citep{GaiaCollaboration2018}. For each galaxy we produced a sky-subtracted, flux-calibrated mosaicked datacube, with a spaxel size of $0.2\arcsec$. Masks for foreground stars were generated based on a combination of the \textit{Gaia} point-source positions and identification of the rest-frame Ca \textsc{ii} triplet absorption features at $8498$, $8542$, and $8662$~\AA. The Ca \textsc{ii} triplet absorption helped to avoid masking compact \hii\ regions and galactic nuclei, which may appear as point sources in \textit{Gaia}. 

The effective R-band PSF of the MUSE data, as measured in our data as detailed in \cite{Emsellem2021}, varies from $0.4\arcsec$ (only achieved in wide-field adaptive optics mode) to $1.2\arcsec$. 
The median PSF for each mosaic is provided in Table~\ref{tab:tab1}, together with the offsets to the largest (max) and smallest (min) PSF measured among the MUSE pointings that constitute the final mosaic. Our median PSF corresponds to an average resolution across the sample of ${\sim}50$~pc.

Throughout the paper, radial distances are deprojected assuming the inclinations and positions angles derived by \cite{Lang2020} from kinematic analysis of molecular gas data, or analysis of near-IR imaging when not available or when the fit to the velocity field was deemed unreliable\footnote{Photometric inclinations are used for IC~5332, NGC~1300, NGC~1365, NGC~1433, NGC~1512, NGC~1672, and NGC~3351.} (see Table~\ref{tab:tab1}).

\subsection{Spectral fitting}
\label{sec:DAP}

The mosaicked cubes are processed by a custom-built data analysis pipeline \citep{Emsellem2021}, based on the public \textsc{gist} (Galaxy IFU Spectroscopy Tool; \citealt{Bittner2019}) software package, and modified by F.~Belfiore and I.~Pessa for obtaining maps of stellar kinematics, stellar population properties (including stellar mass surface density), and emission-line fluxes and kinematics. The pipeline makes use of the penalised pixel fitting python package (\textsc{ppxf}; \citealt{cappellari2004, Cappellari2017}), in three subsequent steps: (i) stellar kinematics extraction, (ii) stellar population analysis, and (iii) determination of emission-line parameters. 

The stellar kinematics and stellar population steps were performed on binned spectra to increase the continuum signal-to-noise ratio (S/N) over that of single spaxels. In particular, we made use of Voronoi binning, as implemented in the python package \textsc{vorbin} \citep{Cappellari2003}, to reach a continuum S/N of 35 in the wavelength range 5300$-$5500 \AA. To fit the stellar continuum we used E-MILES simple stellar population (SSP) models \citep{Vazdekis2016} generated with a \cite{Chabrier2003} initial mass function, BaSTI isochrones \citep{Pietrinferni2004}, eight ages ($0.15{-}14$~Gyr), and four metallicities (${\rm [Z/H]} = [-1.5, -0.35, 0.06, 0.4]$) to determine the stellar kinematics. A slightly larger grid of 78 templates was used for the stellar population analysis.
We fitted the wavelength range $4850{-}7000$~\AA\ in order to avoid strong sky residuals in the redder part of the MUSE wavelength range. The templates were convolved to the MUSE spectral resolution, as derived by \cite{Bacon2017}. 

For the emission-line analysis, we performed an additional, simultaneous fit of stellar continuum and line emission. We also performed this fitting stage with \textsc{ppxf}, exploiting the extensive support for emission-line fitting available within that package, by treating emission lines as additional Gaussian templates. The emission-line analysis was performed on individual spaxels to maximise the resolution of the emission-line maps. During this step, the kinematics of the stellar continuum was fixed to that determined from the binned spectra analysis.  All fluxes and maps presented in this work were corrected for foreground Galactic extinction using the \cite{Cardelli1989} extinction law and the $E(B-V)$ values from \cite{Schlafly2011}.
We reach a typical $3\sigma$ H$\alpha$ surface brightness (\SHa) sensitivity of $1.5 \times 10^{-17}$ erg~s$^{-1}$ cm$^{-2}$ arcsec$^{-2}$ in $0.2\arcsec$ spaxels. More details on the data analysis and quality assessment can be found in \cite{Emsellem2021}.

\subsection{Morphological definitions of DIG and \hii\ regions }
\label{sec:HIIphot}

\hii\ regions present a complex challenge for photometry. Nebulae show irregular morphologies, including shells and arcs, and do not have well-defined boundaries. Sophisticated algorithms have been developed for the task of isolating bright regions from the diffuse background and defining their boundaries. The spatial resolution of the PHANGS--MUSE data (median $50$~pc) is in general not sufficient to resolve the internal structure and size of typical \hii\ regions (${\sim}1{-}100$~pc), except for the most extended, most luminous giant nebulae with size ${>}100$~pc. For comparison, 30~Doradus in the LMC, the most luminous giant \hii\ in the Local Group, is ${\sim}400$~pc in size \citep{Kennicutt1984}. 

In this work we employed a pragmatic definition of \hii\ regions as bright regions with relatively sharp boundaries at the ${\sim}50$~pc resolution of our dataset. We defined \hii\ region masks by running the IDL tool \textsc{HIIphot} \citep{Thilker2000} in conjunction with a point source finder on PSF-homogenised \ha\ line maps. \textsc{HIIphot} is optimised for photometry of extended regions, and separates nebulae from the surrounding DIG by generating seed regions around bright peaks, and then growing them until a termination criterion determined by the spatial gradient of the \SHa\ profile is reached. By visually inspecting the resulting \hii\ region masks we determined that \textsc{HIIphot} was systematically missing a large population of compact point sources, corresponding to unresolved \hii\ regions. In order to obtain a more complete catalogue, we generated masks for these regions using the point-source finder \textsc{DAOStarFinder}, provided by the python \textsc{photutils} module. Our final segmentation maps consists of the union of the \textsc{HIIphot} and \textsc{DAOStarFinder} nebulae masks, and contains ${\sim}1200$ nebulae per galaxy on average. The segmentation maps are shown for all the galaxies in the survey in Appendix~\ref{app:atlas}, Figs.~\ref{HII_region_masks}--\ref{HII_region_masks2}.

The DIG is defined as line emission originating from outside the area covered by the masks used to identify ionised nebulae. Vice versa, the \hii\ region fluxes are taken as simply the flux within the \hii\ region masks, without any correction for the local DIG background. The final catalogue of nebulae is dominated by number by bona fide \hii\ regions, but includes contamination from supernovae remnants, and planetary nebulae (PNe). A detailed description of the catalogue, and further discussion of algorithmic choices made to define \hii\ regions will be presented in a future paper. In the rest of the section we address some of the main limitations, related to blending of nebulae near large complexes, faint nebulae merging with the background, and the difference in kinematics between DIG and \hii\ regions.

Blending of \hii\ regions in crowded regions may affect our ability to determine number counts of \hii\ regions in luminosity bins (i.e. determine the \hii\ region luminosity function, \citealt{Santoro2021}), but should have a small impact on separating \hii\ regions and DIG, since in such regions we expect most of the area to be dominated by \hii\ regions. It should also be noted that defining borders of \hii\ regions in the case of large complexes is an intrinsically ill-defined problem, as the gas may not be easily associated with a unique source of ionising photons. In our analysis, the size of \hii\ regions is set by the termination gradient value we employed in \textsc{HIIphot}, which is driven by the noise level of the data and visual expert assessment of the performance of the algorithm, rather than by physical considerations. Choosing a different termination gradient would extend the \hii\ region masks into the surrounding area, which we have classified as DIG. We experimented with different values of the termination gradient, and find that the line ratios in \hii\ regions are unaffected, while some of the brighter DIG regions get included into the \hii\ region masks. 

A more crucial limitation is the blending of faint \hii\ regions into the diffuse background. The completeness limits of our \hii\ region catalogue are estimated by \cite{Santoro2021} to be in the range $\log(L_\mathrm{H\alpha}/\mathrm{erg~s}^{-1}) \sim 36{-}37$, depending on the distance of the galaxy. \hii\ regions fainter than $\log(L_\mathrm{H\alpha}/\mathrm{erg~s}^{-1}) \sim 36$ have not been studied extensively in external galaxies. In Local Group galaxies, such faint \hii\ regions are found to only contribute marginally to the total \ha\ flux (${<}10$\%, \citealt{Kennicutt1989a} for M33 and the Magellanic Clouds, although M31 may be an exception; see \citealt{Azimlu2011}). Higher-resolution data (provided e.g. by \textit{Hubble} Space Telescope narrow-band imaging) would be necessary to estimate the contribution from faint, undetected \hii\ regions to the DIG.

Planetary nebulae are bright in \oiii\ but comparatively faint in \ha, and are thus not efficiently detected by our algorithm. In \cite{Scheuermann2022} the authors perform a detailed study of the planetary nebula luminosity function (PNLF) in the PHANGS-MUSE galaxy sample, applying a point source detection algorithms to an \oiii\ flux map generated from our MUSE data to detect PNe. Comparing with their catalog of PNe, we find that only 22\% are detected in \ha\, and are therefore identified by our algorithm as nebulae. The remaining PNe identified in \oiii\ are merged into to our DIG mask. We can compute the fractional contribution of PNe to the total DIG flux by integrating the \oiii\ PNLF within suitable limits and normalising it according to the total number of detected PNe in our galaxies, taken from \cite{Scheuermann2022}. If we adopt the PNLF parametrisation of \cite{Ciardullo1989} and integrate it from the bright magnitude cutoff down to sources 8 magnitudes fainter, we find that the total \oiii\ flux contributed by PNe is on average 1\% of the \oiii\ flux observed in our DIG masks. The contribution of PNe to the flux in other emission lines considered in this work, including \ha, will be even smaller. We therefore conclude that, even though our DIG fluxes certainly include a contribution from PNe, this contribution is negligible. Our estimate is in agreement with \cite{Yan2012}, who perform a similar computation, and find that PNe fall short by about 1.5 dex in explaining the observed \oiii\ emission in early-type galaxies.

Detailed decomposition of line profiles may offer a promising tool to distinguish DIG from \hii\ regions, as the DIG is generally found to lie in a thicker, and more slowly rotating structure than the star-forming disc \citep{Fraternali2004, Heald2006, Boettcher2017, Levy2018, DenBrok2020}. However, the rotational velocity lag between DIG and \hii\ regions is of the order of a few tens of $\rm km~s^{-1}$, which is not large enough to be easily detectable at the average spectral resolution of the MUSE data (spectral resolution of $\sim 115~ \mathrm{km~s^{-1}}$ at the wavelength of the \ha\ line). In this paper, therefore, we do not use kinematic information to define the DIG.

\begin{figure*}
        \centering
        \includegraphics[width=0.98\textwidth, trim=0 20 0 0, clip]{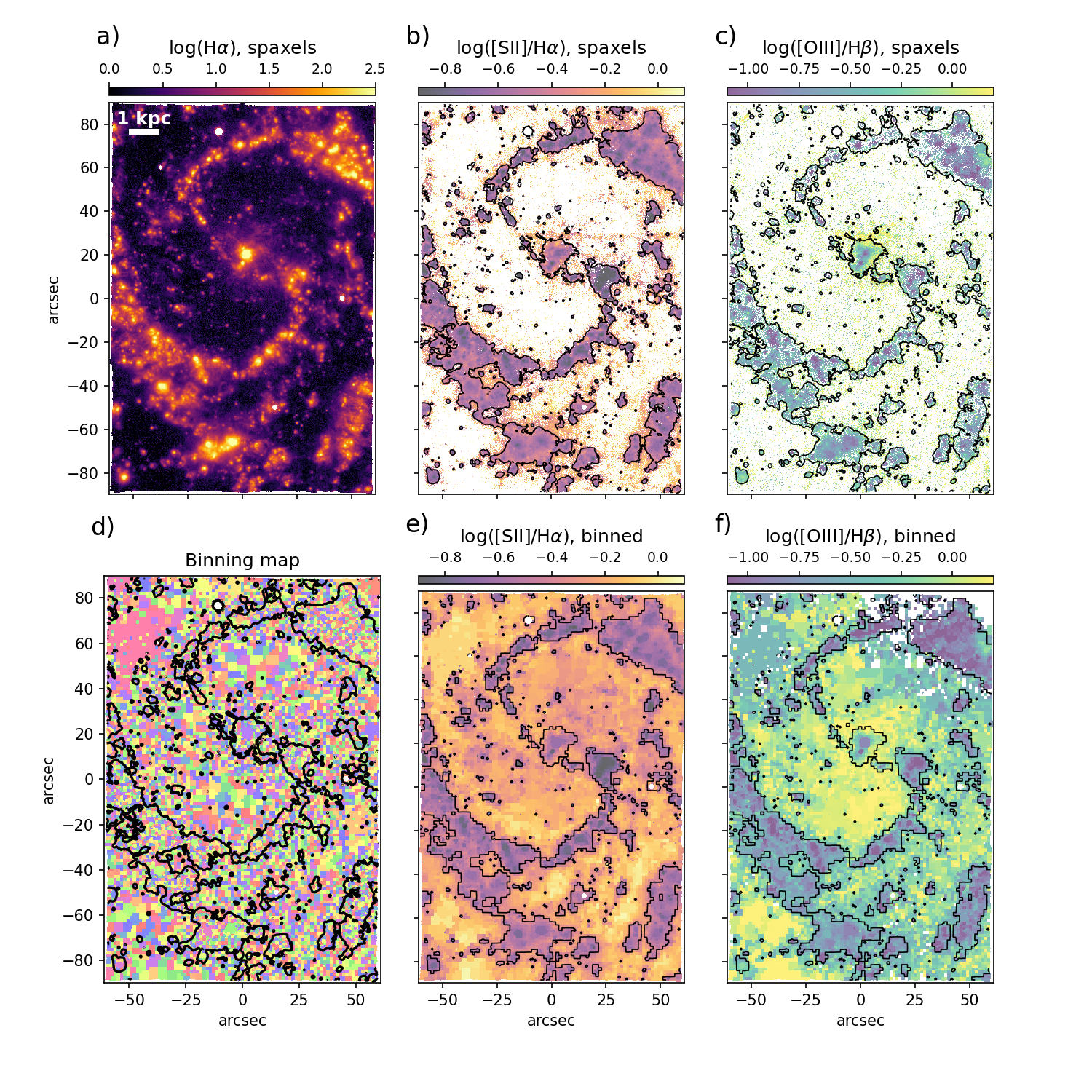}
        \caption{{Maps of the \ha\ flux, binning scheme, and key line ratios for individual spaxels and after Voronoi binning in NGC~4535}. a) Map of $\log(\text{\ha} / 10^{-17} \ {\rm erg~s}^{-1} \ {\rm cm}^{-2} \ {\rm arcsec}^{-2}$), at the native 0.2$''$ single-spaxel resolution. The circular areas that appear white in this and other panels are locations of masked foreground stars. 
        b) Map of log(\sii/\ha) at single-spaxel resolution. White areas have ${\rm S/N} < 3$ for the relevant emission lines. The black contour corresponds to $\Sigma_{\rm H\alpha} = 8 \times 10^{-17}$ erg~s$^{-1}$ cm$^{-2}$ arcsec$^{-2}$ or about a $16\sigma$ detection for H$\alpha$. c) Same as b) but for the log(\oiii/\hb) ratio. d) Representation of the Voronoi binning scheme adopted to recover the low-surface-brightness line emission in the DIG, as described in the text. Each bin is assigned a random colour. e) Map of log(\sii/\ha) using the binned data. f) Same as e) but for the log(\oiii/\hb) ratio.}
        \label{fig_example}
\end{figure*}

\subsection{Spatial binning optimised for low-surface-brightness gas} 
\label{sec:binning}

At the depth of our data, \ha\ is detected throughout a large fraction of the surveyed area (H$\alpha$ is detected at the $3\sigma$ level in 95\% of the $0.2\arcsec$ spaxels in the MUSE observations of our galaxies, \citealt{Emsellem2021}), but even the strongest metal lines, such as \sii\ and \oiii, fall below the detection limit in low-surface-brightness regions. This problem can be better appreciated by inspecting Fig.~\ref{fig_example} (panels a--c), where we show maps of \ha\ flux, log(\sii/\ha) and log(\oiii/\hb) for the galaxy NGC~4535. The black contour corresponds to $\SHa = 8 \times 10^{-17}$ erg~s$^{-1}$ cm$^{-2}$ arcsec$^{-2}$ or about a $16\sigma$ detection for \ha\ in $0.2\arcsec$ spaxels. This surface-brightness level was chosen to correspond roughly to the lower 10$^{\rm th}$ percentile  of \SHa\ for \hii\ regions in this galaxy, and therefore constitutes a useful visual delimiter between bright areas associated with catalogued nebulae and the DIG. In Fig.~\ref{fig_example}, areas with ${\rm S/N} < 3$ in the relevant emission lines are masked out. The figure readily demonstrates the impossibility of deriving representative line ratios in the DIG using individual spaxels.

Our binning choice in the DIG is motivated by the need to obtain a minimum average detection for the metal emission lines of interest. Assuming line ratios typical of the DIG of $\log(\sii/\ha) \sim {-}0.5$ (see also Sect.~\ref{sec:low_ion_data}), and $\log(\oiii/\hb) = {-}0.5$ (see Sect.~\ref{sec:high-ion}), we conclude that in order to achieve a ${\sim}5\sigma$ detection for \oiii\ (and consequently ${\sim}10\sigma$ for $\sii\lambda6717$), one needs an \ha\ S/N of~60. The median \ha\ S/N of single spaxels in the DIG regime is around~9, implying we need to increase the S/N by a mean factor of~seven. We achieve this goal by performing the following procedure. First, we make use of the segmentation map defining ionised nebulae, and treat each nebula as one bin. These regions are not binned further, in order to avoid mixing of flux between \hii\ regions and DIG. Moreover, \hii\ regions are intrinsically bright and need no further binning to ensure detection of the lines of interest.
The DIG area outside the nebulae masks is re-gridded to spaxels of $1.4\arcsec$ (i.e. 7~by~7 native spaxels), in order to increase the S/N by a factor of rougly~seven. The \ha\ S/N in these larger pixels is calculated based on error propagation from the original spaxel-level \ha\ map, since the error vectors in each spaxel are nearly statistically independent \citep{Weilbacher2020a, Emsellem2021}. 

These square $1.4'' \times 1.4''$ bins are used as a starting point for a Voronoi binning procedure, following \cite{Cappellari2003}, which groups together the $1.4'' \times 1.4''$ pixels to reach a target \ha\ S/N of~60. This step is necessary to generate larger bins in regions of lower than average surface brightness in the DIG, and is especially important for galaxies that show large regions of faint line emission (e.g. NGC~1300, NGC~1433, NGC~1512). 
We could have applied the Voronoi binning to the original spaxels, but we found that S/N levels for individual spaxels are not always reliable, leading to bins with unnecessarily complicated small-scale (smaller than the PSF) morphologies. 

We perform a new run of the spectral fitting algorithm to derive the stellar kinematics and emission-line properties on the new bins. The resulting maps for log(\sii/\ha) and log(\oiii/\hb) are shown in Fig.~\ref{fig_example} (panels~e and~f). The figure demonstrates the power of our approach in recovering emission-line ratios in the low-surface-brightness regime. For the remainder of this paper we use these binned emission-line maps as a basis for further analysis of line emission from the DIG. This effectively lowers the spatial resolution of the DIG maps to the size of the average Voronoi bin (${\sim}60$ times the original $0.2\arcsec$ MUSE spaxels), which corresponds to an average scale of $120$~pc across our sample. We note, however, that the higher resolution of the original dataset is crucial to the identification of \hii\ regions and their separation from the DIG.

\section{General characteristics of the DIG}
\label{sec:gen_cat}
 
In this section, we discuss the general characteristics of DIG and \hii\ regions obtained according to our masks derived in Sect.~\ref{sec:HIIphot}. In particular, we discuss key observables that have been used in the literature to distinguish DIG from \hii\ regions: \SHa\ \citep{Zhang2017}, EW(H$\alpha$) \citep[e.g.][]{Lacerda2018}, and the location in the BPT diagram. 
The quantitative conclusions of this section are somewhat dependent on  resolution, and will necessarily differ when galaxies are observed at lower spatial resolution \citep[e.g.][]{Poetrodjojo2019}.

\subsection{\ha\ emission in the DIG and \hii\ regions}
\label{sec:ha_in_DIG}

\begin{figure*} 
        \centering
        \includegraphics[width=1.\textwidth, trim=20 0 20 0, clip]{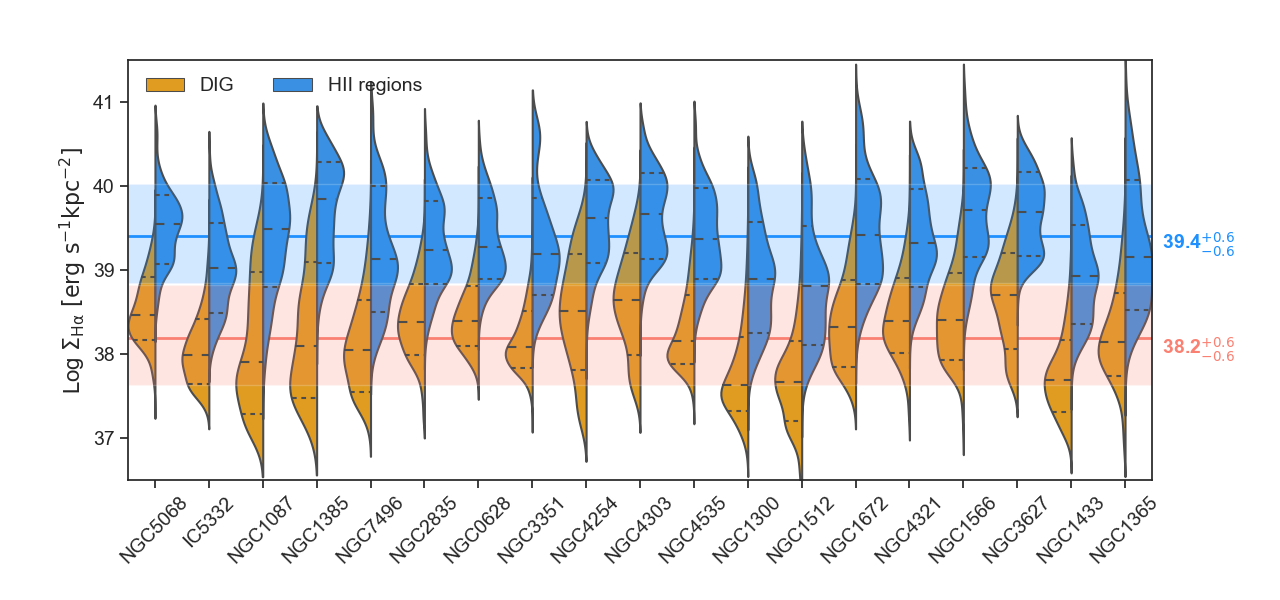}
        \caption{{$\bm\SHa$ in \hii\ regions (in blue) and DIG (in orange) in the PHANGS--MUSE sample ordered by increasing stellar mass from left to right.} DIG and \hii\ regions are defined according to a morphological criterion (using \textsc{HIIphot}; see Sect.~\ref{sec:HIIphot}). The 16$^{\rm th}$, 50$^{\rm th}$, and 84$^{\rm th}$ percentiles of the distributions are shown with dashed lines on each violin. The blue and orange horizontal lines (and associated colour-shaded areas) represent the 50$^{\rm th}$ (and the 16$^{\rm th}$ and 84$^{\rm th}$) percentiles of the \SHa\ distribution for \hii\ regions and DIG across all galaxies. }
        \label{fig_violin_ha}
\end{figure*}

In our sample of galaxies, the DIG defined according to our masks shows variations of ${\sim}1.2$~dex in \ha\ surface brightness, $\log (\Sigma_\mathrm{H\alpha\,DIG} / \mathrm{erg~s^{-1}~kpc^{-2}}) = 38.2\pm0.6$ (where the range shows the 16$^{\rm th}$ and 84$^{\rm th}$ percentiles). In Fig.~\ref{fig_violin_ha}, we show histograms of the \SHa\ distribution in the DIG and \hii\ regions in our sample of galaxies, ordered by increasing stellar mass from left to right. We observe a wide variety in the resulting distributions. Because of our source extraction algorithm, the DIG is effectively defined in terms of its contrast in \SHa\ with respect to clumpy emission, identified as \hii\ regions. As a consequence, while the average \SHa\ of DIG and \hii\ regions varies across galaxies, the average contrast in \SHa\ between DIG and \hii\ region is rather stable at ${\sim}1$~dex. A few galaxies show tails of fainter DIG, in particular NGC~1087 and NGC~1385 show a tail of faint diffuse gas corresponding to the extreme galactic outskirts (galactocentric radii ${>}1 R_{25}$, only covered in these two galaxies and in NGC~4254). For the high-mass galaxies NGC~1300, NGC~1512, and NGC~1433, on the other hand, the faintest regions in \ha\ correspond to the central regions of high mass surface density.

We measure the fraction of the \ha\ flux associated with the DIG ($f_{DIG}$) by summing the \ha\ flux within the DIG masks and correcting this flux for dust extinction using the Balmer decrement. 
 The resulting $f_{DIG}$ is listed in Table \ref{tab:tab1}, and ranges between 20\% and 55\%, with a median value of 37\%. These numbers are in general agreement with previous estimates obtained from \ha\ narrow-band imaging \citep{Oey2007}, and also with the DIG fractions measured independently by \cite{Chevance2020} (median diffuse \ha\ fraction of 34\%) for a subset of eight PHANGS--MUSE galaxies, following the Fourier-filtering methodology of \cite{Hygate2019}.  Our estimate depends on the choice of termination criterion for defining \hii\ region masks. We investigated the effect of changing the termination gradient from our fiducial value to either steeper (resulting in smaller regions) or shallower gradients (resulting in larger regions). These cases are meant to reproduce the range of termination gradient values considered by the authors of \textsc{HIIphot} \citep{Thilker2000} and bracket the range employed in the study of nearby galaxies by \cite{Oey2007}\footnote{In emission measure units our termination gradients correspond to 10 EM $\rm pc^{-1}$ for our steepest and 1.5 EM $\rm pc^{-1}$ for our shallowest, with 5 EM $\rm pc^{-1}$ being our fiducial value}. We find that the median value of  $f_\mathrm{DIG}$ varies from 43\% in the case of steep termination gradients to 22\% for the shallowest one.

The brightest DIG emission is located near bright \hii\ regions in all our galaxies, as can be qualitatively appreciated from  Figs.~\ref{HII_region_masks} and~\ref{HII_region_masks2}. In particular, for galaxies with well-defined spiral arms, the DIG appears brighter in \ha\ near the arms than in the inter-arm regions.
\ha\ emission falls off in intensity with distance from \hii\ regions but otherwise shows a smooth morphology, except in the two galaxies with the lowest masses and smaller distances, NGC~5068 and IC~5332, where prominent shells and filaments are visible in diffuse \ha. Overall, the spatial correspondence of DIG and \hii\ region strongly suggests that leaking radiation powers the DIG, an idea that we develop further in Sect.~\ref{sec:DIG_spatial_model}. 

\subsection{The DIG and \texorpdfstring{EW(H$\alpha$)}{EW(Halpha)}}
\label{sec:HOLMES_HA}

\begin{figure*} 
        \includegraphics[width=0.99\textwidth, trim=0 0 0 0, clip]{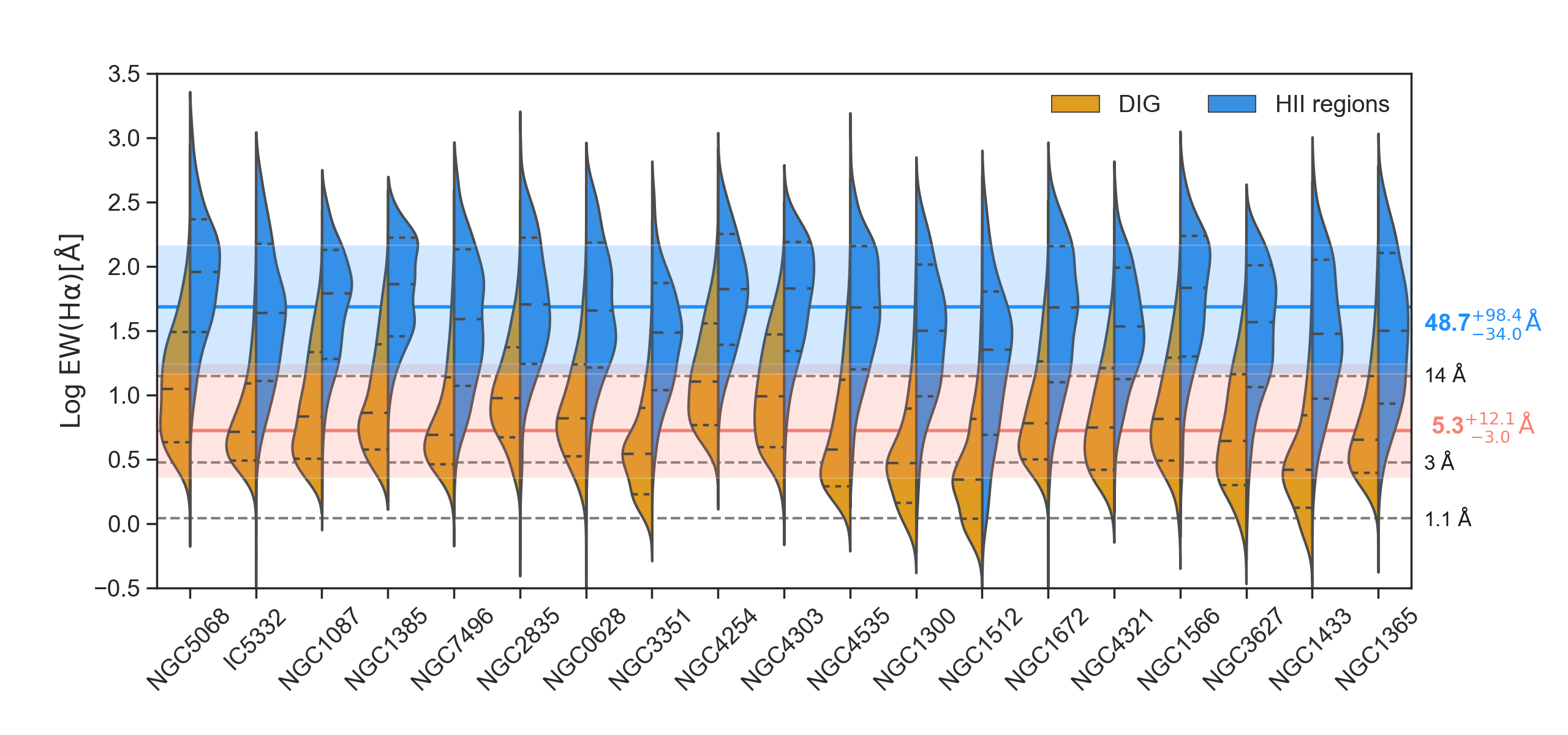}
        \caption{{EW(H$\bm\alpha$) in \hii\ regions (blue) and DIG (orange) in the PHANGS--MUSE sample.} DIG and \hii\ regions are defined according to a morphological criterion (using \textsc{HIIphot}; see Sect.~\ref{sec:HIIphot}). The 16$^{\rm th}$, 50$^{\rm th}$, and 84$^{\rm th}$ percentiles of the distributions are shown with dashed lines on each violin. Dashed horizontal black lines represent demarcations employed in the literature to distinguish \hii\ regions from DIG ($3$~and $14$~\AA) and the value of EW(H$\alpha$) from HOLMES using the \textsc{pegase} models ($1.1$~\AA).  The blue and orange horizontal lines (and associated colour-shaded areas) represent the 50$^{\rm th}$ (16$^{\rm th}$ and 84$^{\rm th}$) percentiles of the EW(\ha)  distribution for \hii\ regions and DIG across all galaxies. }
        \label{fig:violin_ewha}
\end{figure*}

Several studies have discussed the usefulness of EW(\ha) in distinguishing ionisation due to HOLMES from that resulting from star formation or active galactic nuclei (AGN; e.g. \citealt{Stasinska2008, CidFernandes2011}). The value of the EW(\ha) expected from HOLMES remains uncertain, even in the simple case where all the radiation is absorbed `locally'. \cite{CidFernandes2011} compared various population-synthesis codes and found overall agreement to within a factor of two for the number of ionising photons produced by HOLMES. This flux results in $\mathrm{EW}(\ha) = 1.5{-}2.5$~\AA\ for different assumptions regarding metallicity, initial mass function and stellar evolution tracks. 

The number of ionising photons produced by HOLMES declines slowly as a function of age (by a factor of two from $100$~Myr to $10$~Gyr). The EW(\ha) they produce, however, increases with age (by a factor of five from $100$~Myr to $10$~Gyr) because of the decreasing brightness of the stellar continuum, according to the predictions from \textsc{fsps} code \citep{Conroy2009, Byler2019}. In particular, a $10$~Gyr old, solar-metallicity SSP generated with \textsc{fsps} emits $5 \times 10^{40}$ ionising photons ${\rm s}^{-1}~M_{\odot}^{-1}$, corresponding to an $\mathrm{EW}(\ha) \sim 0.5$~\AA.\footnote{We quote the number of ionising photons per unit mass of stars and remnants, which is the same convention for defining stellar masses adopted by the PHANGS--MUSE survey.}
In contrast, the \textsc{pegase} population synthesis predicts a flux of $7 \times 10^{40}$ ionising photons ${\rm s}^{-1}~M_{\odot}^{-1}$ for HOLMES, corresponding to $\mathrm{EW}(\ha) \sim 1.1$~\AA. Given these uncertainties, we take the latter as our fiducial value in this work, as it lies somewhat in the middle of the range between the prediction from \textsc{fsps} and the largest values quoted in \cite{CidFernandes2011}.

In Fig.~\ref{fig:violin_ewha} we show histograms of log(EW(\ha)) of both DIG and \hii\ regions for all the galaxies in the PHANGS--MUSE sample, ordered by stellar mass from left to right. The value of $\mathrm{EW}(\ha) = 3$~\AA, employed in the literature to distinguish \hii\ regions from DIG, does not provide a clean division at the spatial resolution of our data. In fact, the median value of EW(\ha) in the DIG in our sample is $5.3 ^{+12.1}_{-3.0}$~\AA. The value of $14$~\AA, suggested by \cite{Lacerda2018} to select pure \hii\ regions, approximately corresponds to the 84$^{\rm th}$ percentiles of the EW(\ha) distribution of the DIG, and the 16$^{\rm th}$ percentile of the \hii\ regions distribution, therefore providing a reasonable dividing value for our sample. The EW(\ha) distributions are smoother than those of \SHa, simply as a result of the spatially smooth continuum.

In most of the area we classify as DIG, the EW(\ha) is greater than our fiducial value of $1.1$~\AA\ predicted for pure HOLMES. In the central regions of some of the more massive galaxies in our sample, however, we encounter areas of high stellar mass surface density and $\mathrm{EW}(\ha) < 2$~\AA, where HOLMES are predicted to play a major role in keeping the gas ionised. This is particularly evident in, for example, NGC~3351,  NGC~1433, NGC~1300, NGC~1512, and NGC~3627. In some of these galaxies (e.g. NGC~3627) there is evidence for an increase in EW(\ha) towards the centre, potentially associated with the AGN.

\begin{figure*} 
    \centering
        \includegraphics[width=0.85\textwidth, trim=0 0 0 0, clip]{ 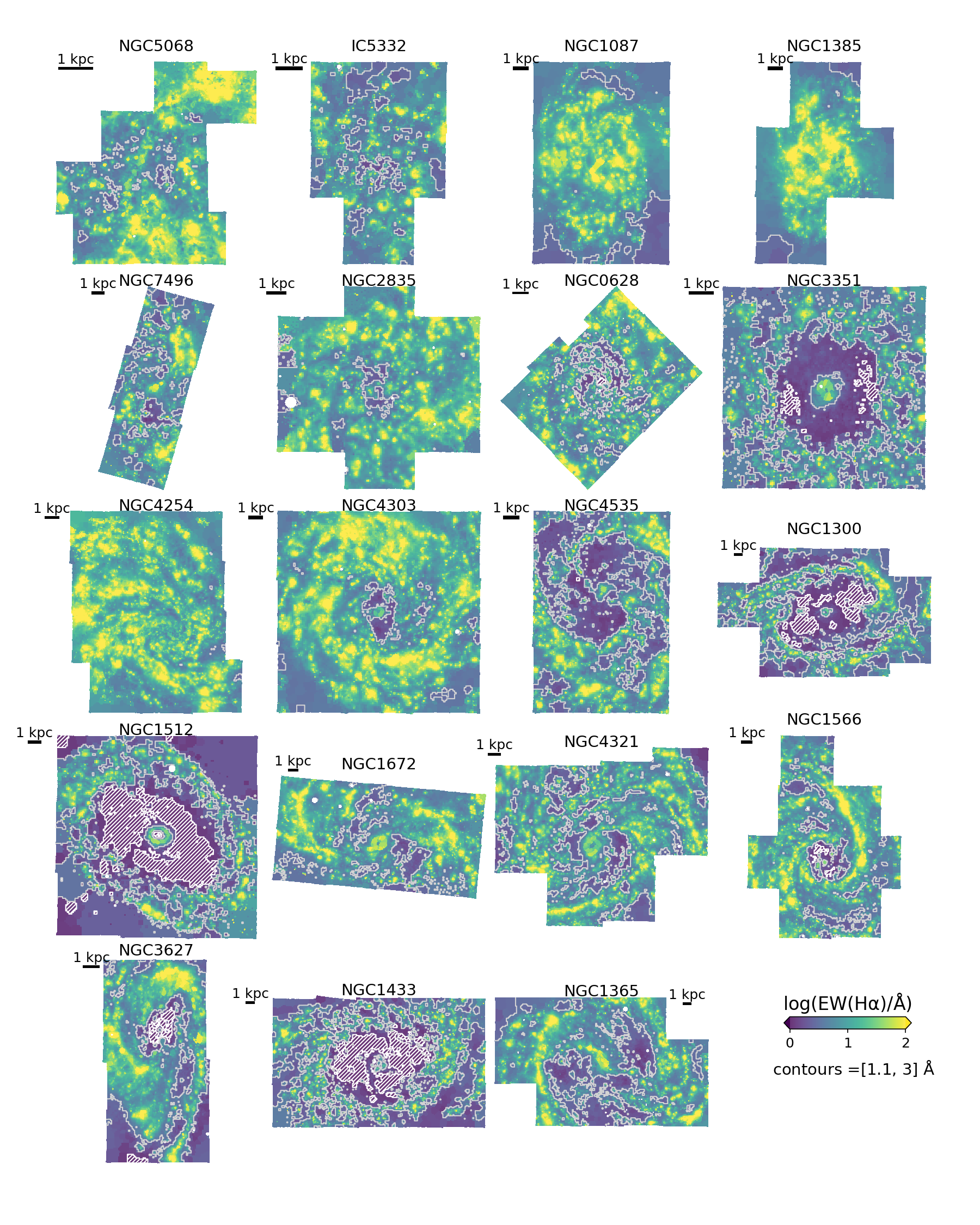}
        \caption{{Maps of EW(H$\bm\alpha$) for all the galaxies in the PHANGS--MUSE sample}. Galaxies are ordered by increasing stellar mass from the top left to the bottom right. The contours correspond to EW(\ha) of $1.1$ (white) and $3$~\AA\ (light grey). For added clarity, regions with $\mathrm{EW}(\ha) < 1.1$~\AA\ are shown with white hatches.}
        \label{fig:ewha_maps}
\end{figure*}

We show EW(\ha) maps for the full sample in Fig.~\ref{fig:ewha_maps}, again ordered by stellar mass along rows from top to bottom. The maps reveal notable differences in the morphology and pervasiveness of regions with $\mathrm{EW}(\ha) < 3$ (grey contour) and $<1.1$~\AA\ (white contour and white hashed for emphasis), respectively. Regions with $\mathrm{EW}(\ha) < 3$~\AA\ start to appear in inter-arm regions (e.g. NGC~7496, NGC~0628) of low-mass galaxies, but become particularly evident in the central regions of higher-mass galaxies swept by a bar (e.g. NGC~3351, NGC~1512, NGC~1433). These are also the only places where we observe extremely low $\mathrm{EW}(\ha) \sim 1$~\AA, compatible with the EW(\ha) predictions for HOLMES at old ages.

\subsection{The DIG in the BPT diagram}

\begin{figure*}
        \centering
        \includegraphics[width=0.85\textwidth, trim=0 0 0 0, clip]{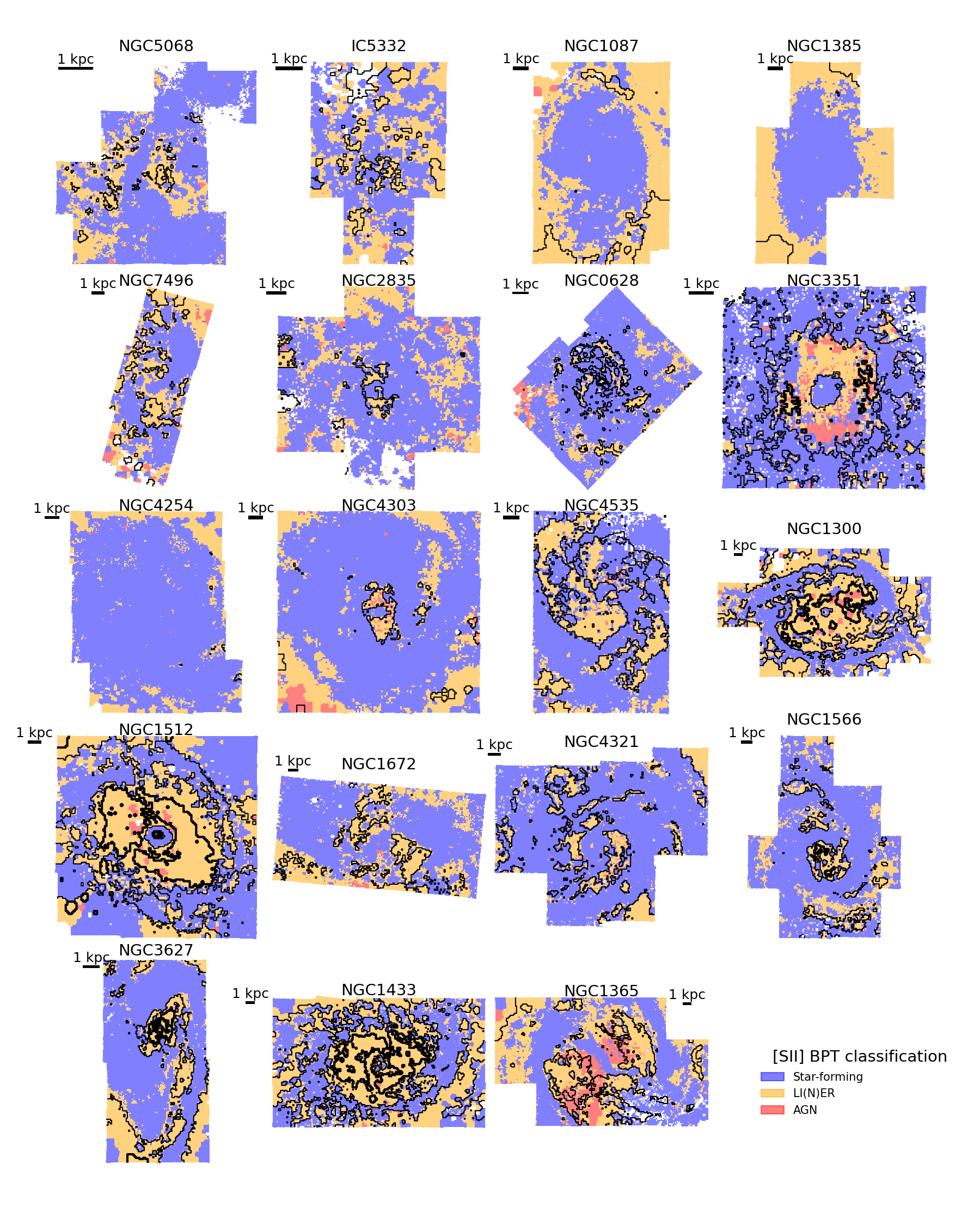}
        \caption{{Maps colour-coded by the position of each region in the \sii\ BPT diagram (blue: star formation; orange: LI(N)ER; red: AGN).} The classification is performed according to the demarcation lines of \cite{Kewley2001} (to classify star-forming regions) and the AGN-LI(N)ER demarcation line of \cite{Kewley2006}. The black contours correspond to EW(\ha) of~$1.1$ and~$3$~\AA\ (same as in Fig.~\ref{fig:ewha_maps}).}
        \label{fig_BPTSII_maps}
\end{figure*}

In Fig.~\ref{fig_BPTSII_maps} we show maps of the galaxies in our sample colour-coded by the position of each region in the \sii\ BPT diagram (\sii/\ha\ versus \oiii/\hb). Similar trends are seen in the \nii\ (\nii/\ha\ versus \oiii/\hb) and \oi\ (\oi/\ha\ versus \oiii/\hb) BPT diagrams. We restrict our discussion to the \sii\ here because of its added ability to discriminate between AGN and LI(N)ER regions with respect to the \nii\ diagram \citep{Kewley2001, Belfiore2016a, Law2020}, and because the \sii\ emission lines are stronger than \oi, therefore allowing classifications for a larger fraction of regions than the \oi\ BPT. We perform only a simple classification into star-forming (blue), LI(N)ER (orange) and AGN (red) regions according to the \cite{Kewley2001} and \cite{Kewley2006} lines. Here we summarise four key points that are evident from these maps.

First, in low-mass galaxies (e.g. NGC~5068, IC~5332, NGC~2835, NGC~1087, NGC~1672, NGC~628), the LI(N)ER-like DIG is found in inter-arm regions and galactic outskirts. These regions correspond to areas of low \SHa, but not necessarily of low EW(\ha).

Second, four galaxies in our sample have AGN signatures in their nuclear regions. NGC~1672 and NGC~1512 show Seyfert-like ionisation in their central regions according to the \sii\ BPT diagram (also confirmed in the \nii/\ha\ versus \oiii/\hb\ plane, and for NGC~1672 by independent evidence from \mbox{X-ray} observations; \citealt{Jenkins2011}). NGC~1566 and NGC~1365 host Type~1 (or intermediate type; see \citealt{Oknyansky2020}) AGN, with broad hydrogen Balmer lines. In NGC~1672, a small region to the south-west of the nucleus shows Seyfert-like line ratios and can be associated with an AGN outflow. NGC~1365 hosts a well-studied biconical AGN ionisation cone \citep{Storchi-Bergmann1991, Kristen1997, Veilleux2003, Venturi2018} to the south-east and north-west of the nucleus, evident as a region of Seyfert-like ionisation in the \sii\ (and all other) BPT diagrams.

Third, NGC~3351,  NGC~1300, NGC~1512, and NGC~1433 host a circum-nuclear star-forming ring, surrounded by a `star-formation desert', with LI(N)ER and Seyfert-like line ratios. These star-formation deserts have very low $\mathrm{EW}(\ha) \sim 1{-}2$~\AA, supporting the idea that their ionisation is largely powered by HOLMES.

Finally, NGC~1566 and NGC~3627 show the signature of kiloparsec-scale, spatially extended LI(N)ER emission in their central regions (equivalent to the typical `central LIER' galaxies observed in kiloparsec-resolution data of nearby galaxies; \citealt{Belfiore2016a}). These central regions also show low $\mathrm{EW}(\ha) < 3$~\AA. This combined evidence supports the idea that line emission in the central regions of these galaxies appears LI(N)ER-like because of the substantial contribution of HOLMES to the local ionisation budget, despite the presence of a Seyfert~1 nucleus in NGC~1566. \footnote{The nucleus of NGC~3627 has been previously classified as `transition object/\linebreak[0]{}Seyfert~2' according to the \cite{Ho1997a} classification. This simply reflects the different classification scheme employed in that work. The line ratios measured by \cite{Ho1997a} for NGC~3627 lead in fact to the same LI(N)ER-like classification we report here.}

In summary, six of our 19 galaxies show extended LI(N)ER regions, four of them in combination with circum-nuclear rings. While the maps in Figs.~\ref{fig:ewha_maps} and~\ref{fig_BPTSII_maps} show large areas with $\mathrm{EW}(\ha) < 3$~\AA\ and LI(N)ER-like ionisation, respectively, these regions contribute a negligible amount of the total \ha\ flux in the galaxy. In particular, on average across our galaxies 88\% of the flux in our DIG masks is contributed by regions that fall in the star-forming section of the \sii\ BPT, while these same regions only occupy 58\% of the DIG on-sky area. While this is clearly different from galaxy to galaxy, it highlights the fact that the DIG with extreme LI(N)ER-like ratio contributes only a small amount of the total \ha\ flux, while occupying a significant fraction of the area.

\section{Spatial model for the propagation of leaking radiation from \texorpdfstring{\hii}{HII} regions} 
\label{sec:DIG_spatial_model}

\subsection{Thin slab model}
\label{sec:model_DIG}

Having determined that HOLMES are subdominant in powering the DIG on galaxy-wide scales, we proceed here to test the association between DIG \ha\ emission and \hii\ regions by building a toy model for the propagation of ionising radiation in the ISM. 

We consider a thin slab model for the propagation of ionising radiation in the DIG, as originally suggested by \cite{Miller1993}, and subsequently explored by \cite{Zurita2002} and \cite{Seon2009}. The model described by \cite{Seon2009} is identical to the one presented in this work. We summarise its derivation in Appendix~\ref{app:spatial_model}. In short, the model considers propagation of ionising photons leaking from \hii\ regions due to both geometric dilution and absorption by a population of DIG clouds. The latter effect is parametrised via an effective absorption coefficient, $k_{0}$. The model predicts that the observed \ha\ surface brightness $\Sigma_{\ha,i}$ of pixel~$i$ is given by
\begin{equation}
   \Sigma_{\ha,i} = \sum_j f_\mathrm{scale} \, \Delta h \, F_{\ha,j}^{\hii} \frac{\mathrm{e}^{-k_{0} r}}{4 \pi r^2}\frac{1-\mathrm{e}^{-k_{0} \Delta r}}{\Delta r}~,
   \label{eq:model}
\end{equation}
where $F_{\ha,j}^{\hii}$ is the \ha\ flux of catalogued \hii\ regions, and the scaling factor, $f_\mathrm{scale} = f_\mathrm{esc} / (1-f_\mathrm{esc})$, where $f_\mathrm{esc}$ is the escape fraction of ionising photons from individual \hii\ regions. $\Delta r$ is the linear size of the pixel considered and $\Delta h$ the height of the thin slab, and $k_{0}$ is the effective absorption coefficient (such that the mean free path of ionisation radiation $\lambda_0 \equiv 1/k_0$).

In order to compare the data with this model, we first subtracted the \ha\ flux due to HOLMES, assuming our fiducial value of the HOLMES ionising photon flux ($7 \times 10^{40}$ ionising photons ${\rm s}^{-1}~M_{\odot}^{-1}$) discussed in Sect.~\ref{sec:HOLMES_HA}, and calculating the ionising flux expected at each location based on the stellar mass surface density of each pixel. This is only an approximate estimate, since it does not account for changes in the ionising flux from HOLMES as a function of age of the stellar population. Nor does it model the propagation of ionising photons from HOLMES. Since we performed a detailed spectral fitting of the MUSE data, it is in principle possible to calculate the HOLMES flux for the specific age and metallicity of the underlying stellar population at each position, but the degeneracies involved in the spectral fitting and uncertainties in model predictions for the post-AGB phase do not currently warrant this level of sophistication. 
 We checked the effect of decreasing or increasing the HOLMES photon flux by a factor of two, and found that this had negligible impact on the conclusions presented in this section. In fact, not performing the subtraction at all only had a small overall impact, causing only marginal shifts in the value of the best-fit parameters.

For each galaxy we assumed $k_{0}$ to be constant across the disc and generated the model \SHa\ via the following sequence of steps. First, we considered the \ha\ flux corrected for extinction (via the Balmer decrement) from each \hii\ region in our catalogue as $F_{\ha,j}^{\hii}$. The flux for each region was taken to be simply the sum of the flux within the region mask. Sources in the \hii\ region catalogue that do not fall within the star-forming demarcation lines of \cite{Kewley2001} in the BPT diagrams of \nii/\ha\ versus \oiii/\hb, \sii/\ha\ versus \oiii/\hb\ and \oi/\ha\ versus \oiii/\hb\  were excluded from this computation since such sources are likely not to be bona fide \hii\ regions, but rather PNe, supernova remnants, or clouds photoionised by the AGN. These sources will clearly also produce ionising radiation, but we excluded them here for the sake of simplicity. Their inclusion does not change any of the results since they contribute only a small fraction (generally a few percent) of the total \ha\ flux in each galaxy. If \hb\ is not detected within the \hii\ region (this only happens for 0.2\% of regions), we assumed no dust extinction. If we do not detect all the metal lines necessary for the BPT classification, we assumed the source is a faint \hii\ region and retain it in the sample (this only affects 2\% of the regions). Different choices for these edge cases are inconsequential to the subsequent results.

For each \hii\ region, we considered its \ha\ luminosity-weighted centre (using the geometric centre does not change any of our conclusions) and computed the deprojected distance to other pixels in the galaxy, taking the position angle and inclination into account. We only considered radiation propagating along the midplane of the galaxy. We used the deprojected distance, $r_{ij}$, to compute the \ha\ emission in the DIG according to Eq.~\ref{eq:model}. In particular, since $f_{\rm scale}$ and $\Delta h$ are degenerate with each other, we computed the radiation field described in Eq.~\ref{eq:model} up to a multiplicative factor. 
 The multiplicative scaling factor was subsequently determined by comparing the model with the data (Sect.~\ref{sec:model_vs_data}).
     
We did not correct the DIG \ha\ emission for dust extinction, therefore effectively assuming no extinction in the diffuse medium. This assumption is not strictly correct as we measure an average $E(B-V) = 0.15$ mag in the DIG, based on the Balmer decrement. We also did not treat the effect of dust in the DIG on the propagation of ionising photons. Radiative transfer models from \cite{Pellegrini2020} demonstrate that this is a reasonable assumption for low-inclination galaxies.  
     
Finally, we convolved the model with the median PSF of the MUSE mosaic and spatially binned it using the same bins as in the observed \ha\ map.

 \begin{figure*} 
        \includegraphics[width=0.99\textwidth, trim=0 0 0 0, clip]{ 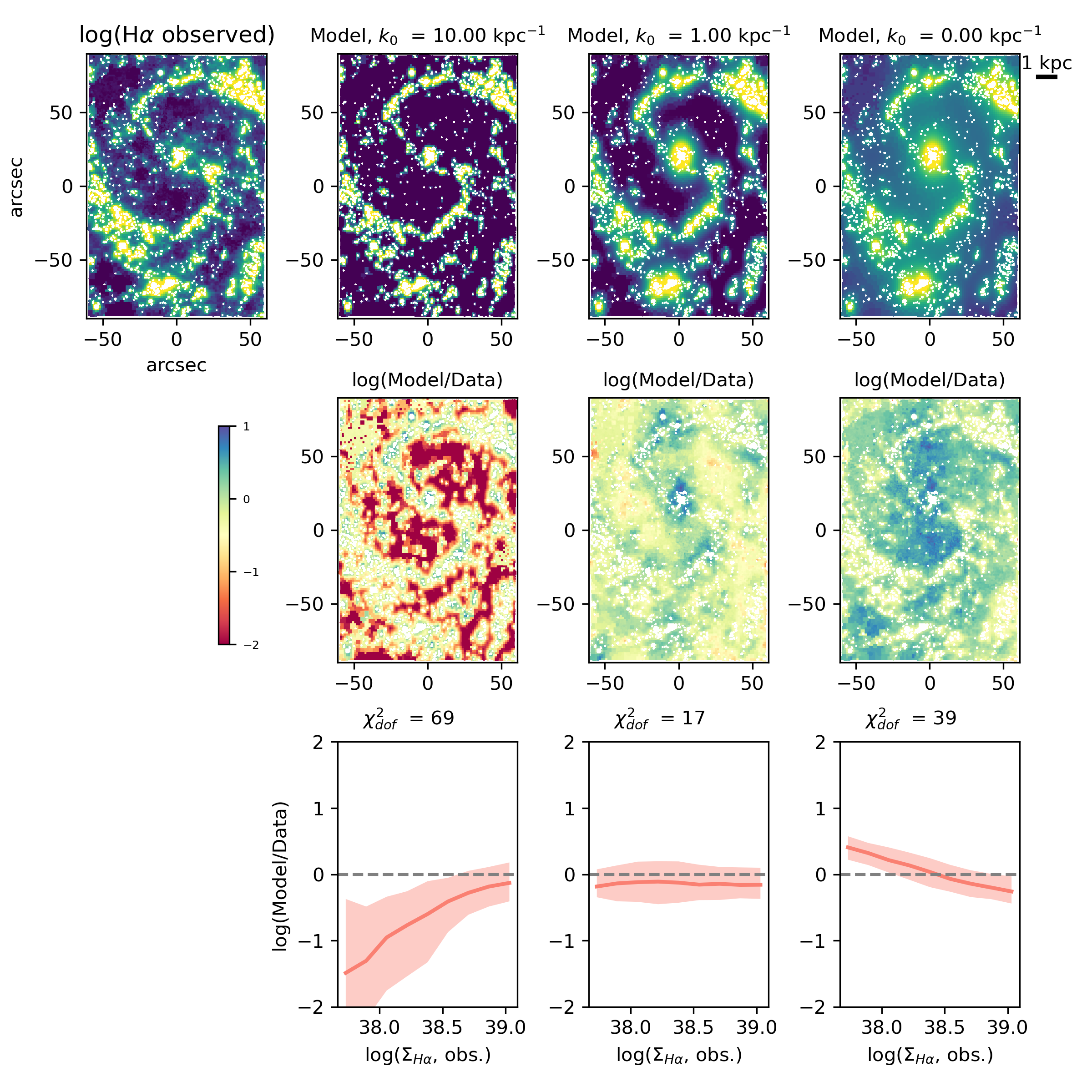}

        \caption{{Example of our DIG modelling and its comparison with the observed $\SHa$ for NGC~4535.} In the top row, we show the observed \ha\ flux map (\hii\ regions are masked and appear white in all the maps) and the model predictions for different values of $k_0$, generated according to Eq.~\ref{eq:model}. In the middle row, we show a spatial comparison of the model and observations, in the form of log(Model/Data). In the bottom row, we show the median relation between log(Model/Data) and $\Sigma_\mathrm{\ha,\,observed}$ with associated scatter (the shaded region corresponds to the range between the 16$\rm ^{th}$ and 84$\rm ^{th}$ percentiles). The $\chi^2_\mathrm{dof}$ for each model is also listed above each panel. The model with $k_0 \sim 1~{\rm kpc}^{-1}$ offers a better representation of the data with respect to models with higher ($k_0 = 10~{\rm kpc}^{-1}$) or lower values ($k_0 = 0~{\rm kpc}^{-1}$, i.e.\ only spherical dilution).}
        \label{fig:thin_slab_model_1}
 \end{figure*}

 \begin{figure*} 
        \includegraphics[width=0.99\textwidth, trim=0 30 0 0, clip]{ 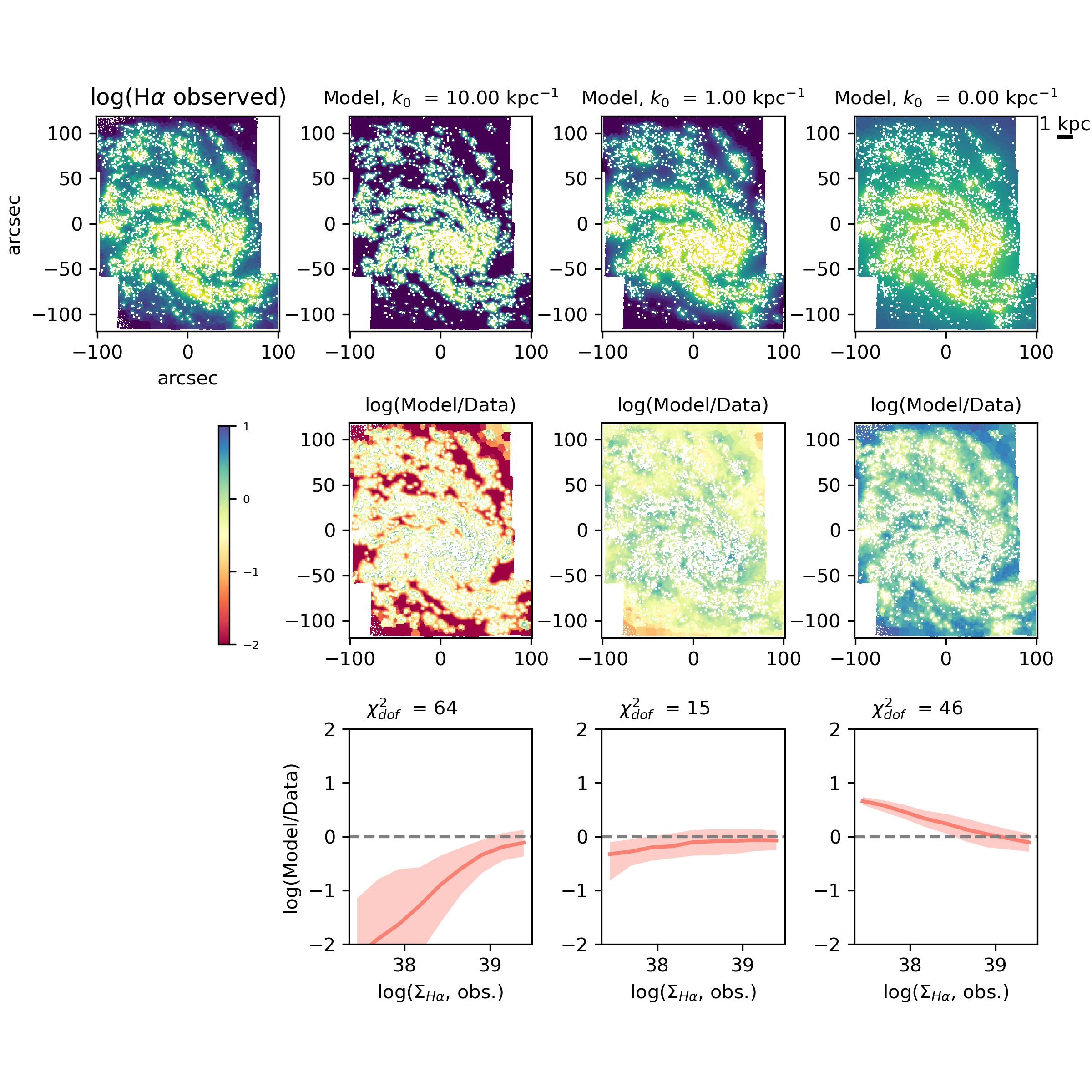}
        \caption{Same as Fig.~\ref{fig:thin_slab_model_1} but for NGC~4254. }
        \label{fig:thin_slab_model_2}
 \end{figure*}

 \subsection{Comparing the observed \ha\ with a leakage model}
 \label{sec:model_vs_data}

Models were computed for nine logarithmically spaced values of $k_{0} = [0.1, 10.0]~{\rm kpc}^{-1}$. We also computed a model with $k_{0}=10^{-4} \sim 0~{\rm kpc}^{-1}$, which corresponds to the case of spherical dilution, since in such a model the mean free path of ionisation radiation $\lambda_0 = 10^4$~kpc is much larger than the physical size of the system (${\sim}10$~kpc). 

 For each galaxy in our sample and each value of $k_{0}$ we compare the model predictions with the observed \ha\ map for the areas classified as DIG (\hii\ regions are masked). Since we only computed the model \ha\ distribution up to a multiplicative scale factor, we need to rescale the model to the observations. We do so by fitting a straight line of the form 
 \begin{equation}
     \Sigma_{\ha,\,\mathrm{observed}, i} = a_0 + a_1 \sum_j F_{\ha, j}^{\hii} \frac{\mathrm{e}^{-k_{0} r_{ij}}}{4 \pi r_{ij}^2}~.
     \label{eq:fit_model_to_data}
 \end{equation}
 According to Eq.~\ref{eq:model}, the slope parameter $a_1$ is proportional to $f_{\rm scale} \, \Delta h$, while $a_0$ should be zero if the model can correctly predict the DIG observations without any additional background.

 In Figs.~\ref{fig:thin_slab_model_1}  and~\ref{fig:thin_slab_model_2} we illustrate the results of this procedure for the two example galaxies, NGC~4535 and NGC~4254. For each galaxy we present the observed \SHa\ map of the DIG (\hii\ regions are masked and appear white) and the model predictions for different values of $k_0$ (top row), maps of the logarithmic difference between model and data (middle row) and examples of the median relations between log(best-fit model/data) and $\Sigma_{\rm H\alpha,\, observed}$. The $\chi^2_\mathrm{dof}$ for each model is also indicated, which is computed by considering an intrinsic scatter of 10\%, added in quadrature to the observed flux errors. 

 Both galaxies manifest similar trends in this comparison. Considering the model with $k_0 = 10~{\rm kpc}^{-1}$, the radiation field does not travel far enough from the \hii\ regions. The observed \SHa\ is higher than that predicted by the model in low \SHa\ regions, explaining the large areas of negative log(model/data). At the other extreme, in the case of spherical dilution ($k_0 = 0$), the model predicts too much diffuse flux with respect to the data. For both example galaxies, the spherical dilution model overestimates the \ha\ flux at low \SHa\ and underestimates it in the vicinity of \hii\ regions (high \SHa). 

 For $k_0 = 1~{\rm kpc}^{-1}$, both NGC~4535 and NGC~4254 exhibit a nearly flat relation between log(model/data) and $\Sigma_{\ha,\,\mathrm{observed}}$. The model predicts roughly the correct amount of \ha\ both in the vicinity of  \hii\ regions and in the diffuse component. While such models still show a relatively high $\chi^2_\mathrm{dof}$, we consider the fit satisfactory given the simplistic assumptions we have employed.

 \begin{figure*} 
        \includegraphics[width=0.99\textwidth, trim=0 0 0 0, clip]{ 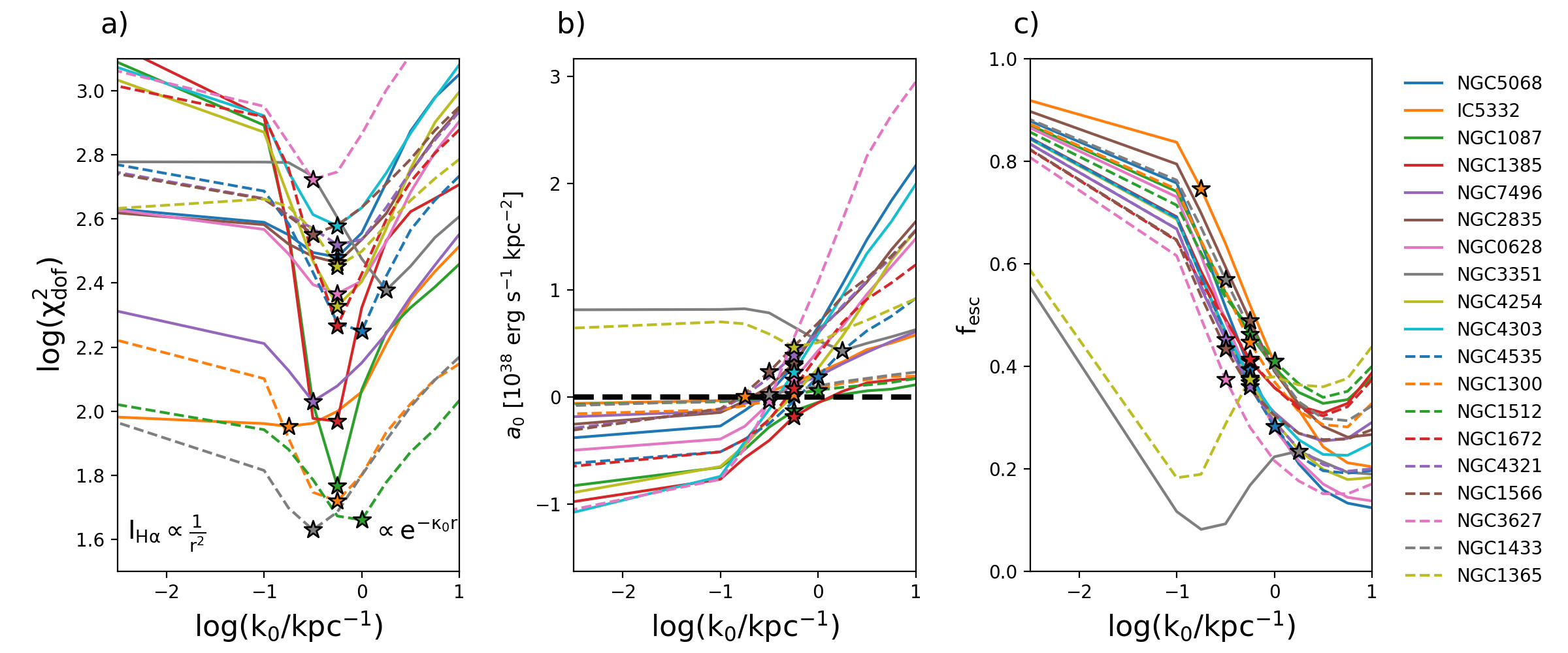}
        \caption{{Summary of best-fit model parameters for the DIG leakage model applied to our entire sample.}
        \textit{a)} Variation in $\chi^2_\mathrm{dof}$ for the DIG spatial models described in the text as a function of the absorption coefficient, $k_0$, for all the galaxies in our sample. Low values of $k_0$ correspond to the case of spherical dilution, while high values correspond to exponential attenuation with scale length $1/k_0$. For each galaxy, a coloured star denotes the model with minimum $\chi^2_\mathrm{dof}$. \textit{b)} Value of the additive offset, $a_0$, defined in Eq.~\ref{eq:fit_model_to_data}, as a function of  $k_0$. According to the simple model described in the text, $a_0 \sim 0$. Star symbols correspond to the $k_0$ values with the lowest $\chi^2_\mathrm{dof}$, which are clustered around $a_0 \sim 0$. \textit{c)} $f_\mathrm{esc}$, the escape fraction of ionising radiation from \hii\ regions as a function of $k_0$ for each galaxy, assuming $\Delta h = 1~ \mathrm{kpc}$. Star symbols correspond to the $k_0$ values with the lowest $\chi^2_\mathrm{dof}$.}
        \label{fig:thin_slab_model_chi2}
 \end{figure*}

 \subsection{Best-fit model parameters}
 \label{sec:best_fit_params}

 We present the variations in the $\chi^2_\mathrm{dof}$ of the fit as a function of $k_0$ for each galaxy in Fig.~\ref{fig:thin_slab_model_chi2}a. A star symbol denotes the position of the model with the lowest $\chi^2_\mathrm{dof}$. The figure demonstrates that for all galaxies we observe a well-defined $\chi^2_\mathrm{dof}$ minimum. For the example galaxies in Figs.~\ref{fig:thin_slab_model_1} and~\ref{fig:thin_slab_model_2}, NGC~4535 and NGC~4254, the lowest $\chi^2_\mathrm{dof}$ corresponds to $\log(k_0/{\rm kpc}^{-1}) = 0.0$ and $\log(k_0/{\rm kpc}^{-1}) = -0.25$, respectively, or mean free paths of $1.0$ and $1.8$~kpc. Across the full sample, the average mean free path of the best-fit model is $1.9$~kpc ($k_0 = 0.52 \ {\rm kpc}^{-1}$). This scale can be compared with the mean nearest neighbour distance between \hii\ regions in our catalogue ($65{-}300$~pc, median 170~pc), and that measured by \citep[][$80{-}190$~pc, median 180~pc]{Chevance2020} for a subsample of eight of our targets, showing that the DIG emission is truly diffuse and occurs on scales much larger then the separation scales of the units of the star-formation cycle\footnote{The nearest-neighbour distance between \hii\ regions can be computed from the parameters given in \cite{Chevance2020} as $ 0.443~ \lambda \sqrt{ \frac{t_{\mathrm{CO}} + t_{\mathrm{star,ref}} }{t_{\mathrm{fb}}+t_{\mathrm{star,ref}}}} $, where $\lambda$ is defined as the characteristic separation between units of the star-formation cycle (which includes both \hii\ regions and giant molecular clouds) and $t_{\mathrm{CO}}$, $t_{\mathrm{star,ref}}$, and $t_{\mathrm{fb}}$ are the durations, respectively, of the CO-emitting phase, the \ha-emitting phase, and the overlap phase between CO and \ha\ emission (dubbed `feedback phase'). The factor of 0.443 refers to the averaging of the nearest neighbour distribution function, as explained in \cite{Kruijssen2019}, Eq. 9.}. We do not find any significant correlations between the best-fit value of $k_0$ and the stellar mass, gas fraction, distance or inclination of the galaxies in our small sample.

 Figure~\ref{fig:thin_slab_model_chi2}b presents the variation in the additive offset $a_0$ as defined in Eq.~\ref{eq:fit_model_to_data}. This parameter should be equal to zero, if the DIG is well described by our modelling framework. We find in Fig.~\ref{fig:thin_slab_model_chi2}b that $a_0$ is close to zero for the models with the lowest $\chi^2_\mathrm{dof}$ (coloured stars). The value of $a_0$ increases from negative to positive with increasing $k_0$, implying that there is a value of $k_0$ for which $a_0=0$ exactly (i.e. it is suitable to reproduce the DIG without any constant background). For NGC~3351 and NGC~1365, such a change in sign of $a_0$ is not observed. Both galaxies have extended regions (the central regions in NGC~3351 and an AGN ionisation cone in NGC~1365; see Sect.~\ref{sec:DIG_BPT}) that are not ionised by leaking radiation from \hii\ regions. 

 Finally, the multiplicative offset $a_1$, also defined in Eq.~\ref{eq:fit_model_to_data}, is related to the product $f_{\rm scale} \, \Delta h$. Measurements of the DIG scale height in our own galaxy or in edge-on spiral galaxies span a range of $0.5{-}1.5$~kpc \citep{Reynolds1991, Ferriere2001, Gaensler2008, Levy2019}. Here we take the median value of 1~kpc obtained by \cite{Levy2019} using a compilation of literature measurements. The thickness $\Delta h$ is given by twice the scale height, but it is reasonable to assume that the \ha\ emission from the far side of the galaxy with respect to the disc midplane is obscured, and therefore not observed. Assuming $\Delta h = 1$ kpc and taking into account a correction for galaxy inclination, we can derive an estimate of the escape fraction $f_\mathrm{esc}$. We show the derived value of $f_\mathrm{esc}$ for each galaxy as a function of $k_0$ in Fig.~\ref{fig:thin_slab_model_chi2}c. For each galaxy the model with the lowest $\chi^2_\mathrm{dof}$ is shown with a coloured star. As expected, for most galaxies a smaller $k_0$ (and therefore a larger mean free path) requires a larger escape fraction to fit the data. For the best-fit models, the median escape fraction is 40\%, with IC~5332 being an outlier with $f_\mathrm{esc} > 0.6$. These values of escape fraction are consistent with those obtained in Sect. \ref{sec:ha_in_DIG}, if one assumes $f_\mathrm{DIG} = f_\mathrm{esc}$.
 
Considering the assumed thickness of the DIG layer, the toy model for a thin disc presented in Sect.~\ref{sec:DIG_spatial_model} may not be a good approximation since the emitting region is not particularly thin, even though the galaxy as a whole is. \cite{Seon2009} compared the thin-slab model also used in this work with a 3D Monte Carlo radiative transfer code, confirming that they lead to similar results for $\Delta h/2 < 1/k_{0}$. 

 In summary, the simple model for the propagation of ionisation radiation that we have developed in this section is capable of reproducing reasonably well the observed \SHa\ in the DIG for a median value of the mean free path of ionising radiation of $1.9$~kpc. While our models do not reproduce the small-scale variations in \SHa\ observed in the data, as reflected by the high best-fit $\chi^2_\mathrm{dof}$, they are nonetheless capable of reproducing the average observed \SHa\ in the DIG from the lowest to the highest observed \ha\ surface brightness. Our conclusions are comparable to those drawn by \cite{Zurita2002} and \cite{Seon2009} for NGC~157 and M~51, respectively. 

The spatial model for the DIG we have developed potentially allows us to calculate the fraction of the \ha\ flux within each \hii\ region spatial mask due to radiation leaked from all other regions in the galaxy. Subtracting this DIG contamination from the \ha\ flux provided in our \hii\ region catalogue would allow for a more accurate estimate of the intrinsic emission of ionising photons for each \hii\ region (however, the contribution from leaking radiation along the line of sight from the region itself would have to be estimated independently). A change in the intrinsic \hii\ region flux would, in turn, result in a change in the DIG model, requiring a recursive approach to achieve convergence. Given the uncertainties in the modelling framework, we do not attempt to follow this approach here. We note, however, that the main effect of these corrections would be to lower the luminosity of \hii\ regions in areas of high spatial clustering (e.g. spiral arms) and therefore larger DIG surface brightness. This may potentially reduce the discrepancies observed in Fig.~\ref{fig:thin_slab_model_1} and~\ref{fig:thin_slab_model_2}, where the best-fit models show too much flux in the vicinity of the spiral arms.

The conclusion from this modelling effort is that the DIG \SHa\ is consistent with photoionisation from LyC radiation leaking from \hii\ regions. Since successful models can be built with $k_0$ being constant throughout the disc, the clumping (or volume filling) factor of photoionised clouds in the DIG is roughly constant within galaxies, and does not change much across our sample. These findings motivate the study of line ratios in the DIG as a function of \SHa\ presented in the next section.

\section{Line ratios in the DIG}
\label{sec:line_ratios}

\subsection{Dependence of line ratios on \SHa\ and metallicity}
\label{sec:lines_intro}

According to the analysis in the previous section, in the DIG the distance between the source of the ionising photons and the absorbing medium is much larger than the typical radii of \hii\ regions. In order to predict the line ratios in a photoionised medium one needs to consider the ratio between the photon number density and the electron density, referred to as the ionisation parameter,~$U_0$, defined as
\begin{equation}
U_0 = \frac{\Phi_i}{n_\mathrm{e} \, c}~,
\end{equation}
where $\Phi_i$ is the number of ionising photons reaching the incident surface of the ionised cloud per unit time and unit area.\footnote{The subscript in $U_0$ highlights the fact that we consider the ionisation parameter at the incident face of the cloud. See \cite{Kewley2019} for a discussion of different ways of defining the ionisation parameter in photoionisation models. In particular, here we assume $\Phi_i = Q_{\rm ion} / (4 \pi r^2)$ where $Q_{\rm ion}$ is the number of ionising photons per unit time emitted by the source and $r$ is the distance between the source and the inner surface of the cloud in case of spherical geometry.}
In the low electron density limit, which always applies to the DIG, the thermal and ionisation structure of a nebula depends only on $U_0$, and not on $\Phi_i$ and $n_\mathrm{e}$ individually. 

The electron density in the DIG cannot be directly measured with our data. Typical density-sensitive line ratios in the optical wavelength range, such as the $\sii\lambda6717 / \sii\lambda6731$ ratio, saturate at their low-density limit for $n_\mathrm{e}$ less than a few tens of $\rm cm^{-3}$ and are therefore unsuitable to measure the density of the DIG. Studies of the diffuse emission in the Milky Way have determined that the typical electron density of DIG clouds is $n_\mathrm{e} \sim 0.1~{\rm cm}^{-3}$ \citep{Reynolds1991, Ferriere2001}, about $2{-}3$~dex lower than typical values for \hii\ regions \citep{Kennicutt1984, Dale2009a, Kewley2019}. Since the typical sizes of \hii\ regions ($10{-}100$~pc) are $1{-}2$~dex smaller than the mean free path of ionising photons in the DIG as determined in Sect.~\ref{sec:best_fit_params}, we predict the ionisation parameter in the DIG to be up to ${\sim}2$~dex lower than in \hii\ regions. Typical values for the ionisation parameter in \hii\ regions lie in the range $\log(U_0) = [-3.5, -2.5]$ \citep{Kewley2019, Mingozzi2020}. We therefore expect the DIG to have $\log(U_0) = [-4.5, -3.5]$.
If the electron density is roughly constant within the DIG layer, \SHa\ is proportional to the ionisation parameter, since we have demonstrated in Sect.~\ref{sec:model_DIG} that \SHa\ traces the diluted ionising radiation field.  

The interpretation of line ratios in the DIG, especially in external galaxies, is complicated by the difficulty of isolating the effects of changes in ionisation parameter from variations in both metallicity and the shape of the ionising continuum. In light of the discussion in the previous section, we assume here that the DIG is mostly ionised by radiation from young stars leaking from \hii\ regions. We discuss in Sect.~\ref{sec:holmes_models} the effect of the harder radiation field from HOLMES. 

Having assumed a fixed input spectrum, we marginalise over the effects of metallicity on the line ratios in the DIG by looking for trends in bins of galactocentric radius. Since galaxies present well-defined metallicity gradients \citep{Moustakas2010, Sanchez2014, Ho2015, Belfiore2017, Mingozzi2020}, and azimuthal variations in chemical abundances in our galaxies are small (${<}0.05$~dex; \citealt{Kreckel2019, Kreckel2020}), we assume that within our radial bins (described in Sect. \ref{sec:low_ion_data}) the DIG is roughly chemically homogeneous.

We study the changes of line ratios in the DIG following the approach of \citet{Zhang2017}, who studied the line ratios in the DIG as a function of both \SHa\ and galactocentric distance using data from the MaNGA survey. Such an approach allows some separation of the effect of ionisation parameters, via \SHa\ as its proxy, from metallicity, assumed to decrease with galactocentric distance. Unlike \citet{Zhang2017}, however, we resolve the mean separation scale between \hii\ regions and can therefore measure the DIG line ratios without being contaminated by flux from \hii\ regions.

\subsection{Low-ionisation line ratios in the DIG}

\begin{figure*} 
        \includegraphics[width=0.99\textwidth, trim=0 0 0 0, clip]{ 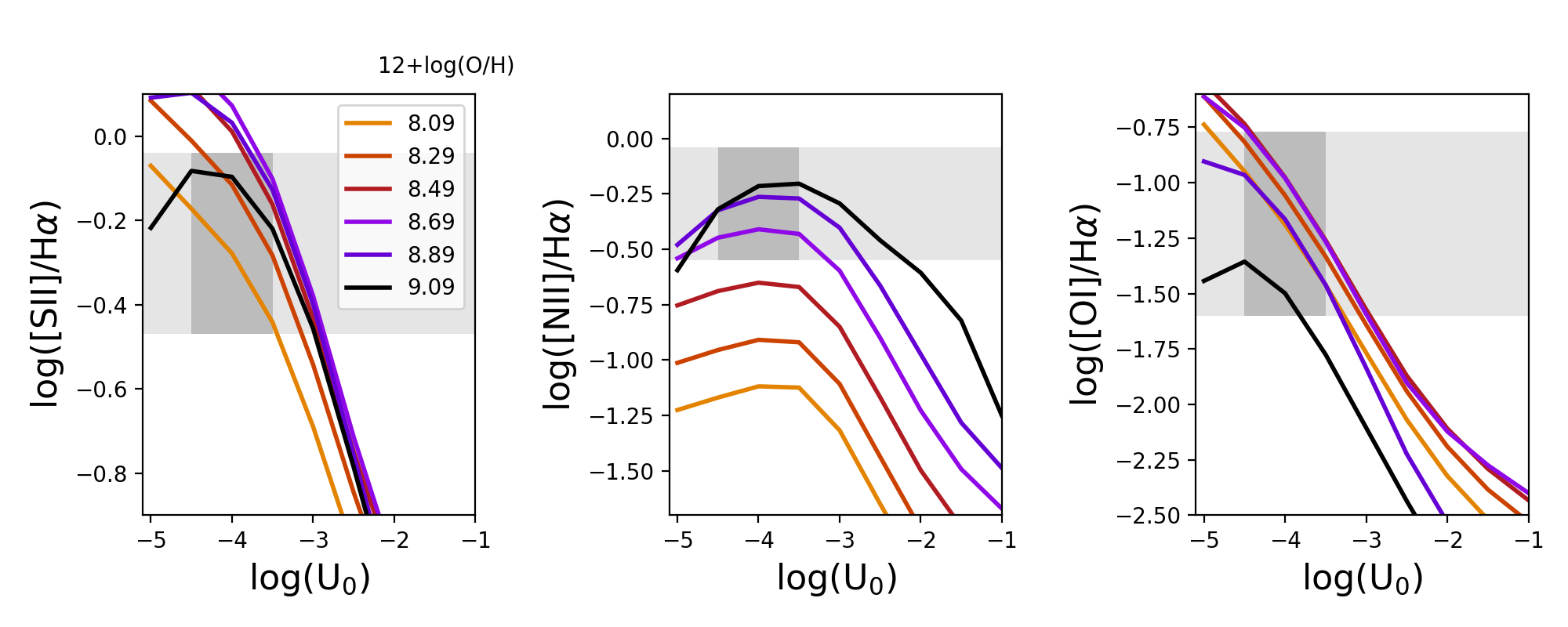}
        \caption{{Model predictions for changes in low-ionisation line ratios with ionisation parameter in the case of leaking radiation.}  Models are computed with \textsc{cloudy} v17.02 and a 2~Myr SSP spectrum generated with \textsc{fsps} as input. The line ratios, from left to right log(\sii/\ha), log(\nii/\ha), and log(\oi/\ha), are plotted as a function of the ionisation parameter at the incident surface of the cloud ($\log(U_0)$). Different colours correspond to different gas-phase metallicities, as indicated in the legend. The light grey band corresponds to the 10$^{\rm th}$ and the 90$^{\rm th}$ percentiles of the observed line ratios in the DIG across all the galaxies in our sample. We further highlight in dark grey the area of $\log(U_0) = [-4.5, -3.5]$, corresponding to the expected range of ionisation parameter values in the DIG. The models span the range of observed line ratios in the DIG for the expected values of $\log(U_0)$ in the case of log(\sii/\ha) and log(\oi/\ha), while they fall somewhat short of reproducing the highest line ratios for log(\nii/\ha).  }
        \label{fig:models_lowion}
\end{figure*}

\begin{figure*} 
        \includegraphics[width=0.99\textwidth, trim=0 0 0 0, clip]{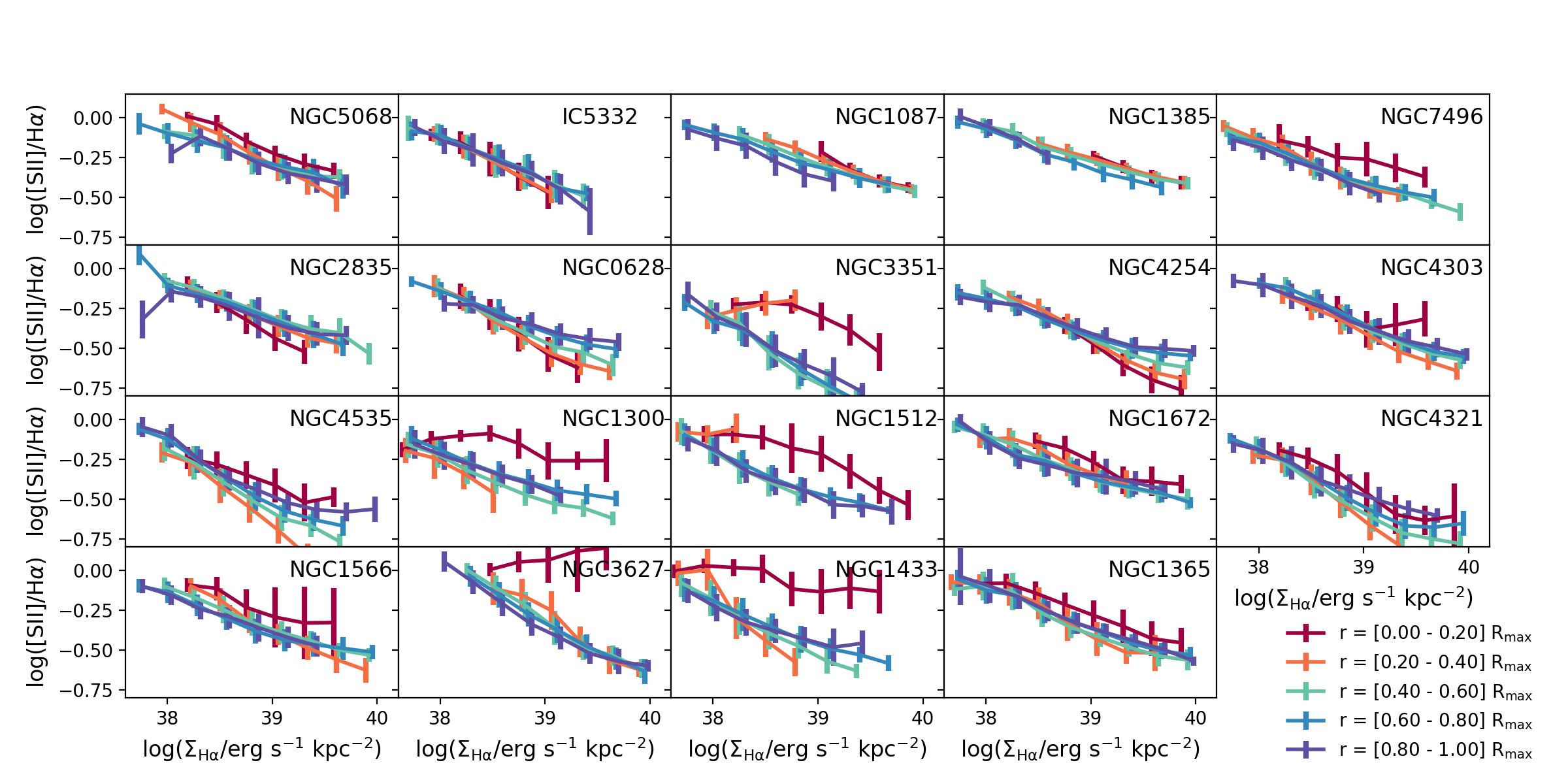}
        \caption{{Log(\sii/H$\bm{\alpha}$) in the DIG as a function of log($\bm{\Sigma_\mathrm{H\alpha} / \textrm{erg~s}^{-1}~\textrm{kpc}^{-2}}$) for different radial bins in individual galaxies in the PHANGS--MUSE sample.} Galaxies are ordered by stellar mass from lowest (top left) to highest (bottom right). Several galaxies (NGC~3351, NGC~1300, NGC~1512, NGC~3627, and NGC~1433) have higher values of this (and other) line ratios in their central radial bin (in red). Aside from this feature, the galaxies in the sample follow a remarkably tight relation between log(\sii/\ha) and \SHa.}
        \label{figS2}
        
        \includegraphics[width=0.99\textwidth, trim=0 0 0 0, clip]{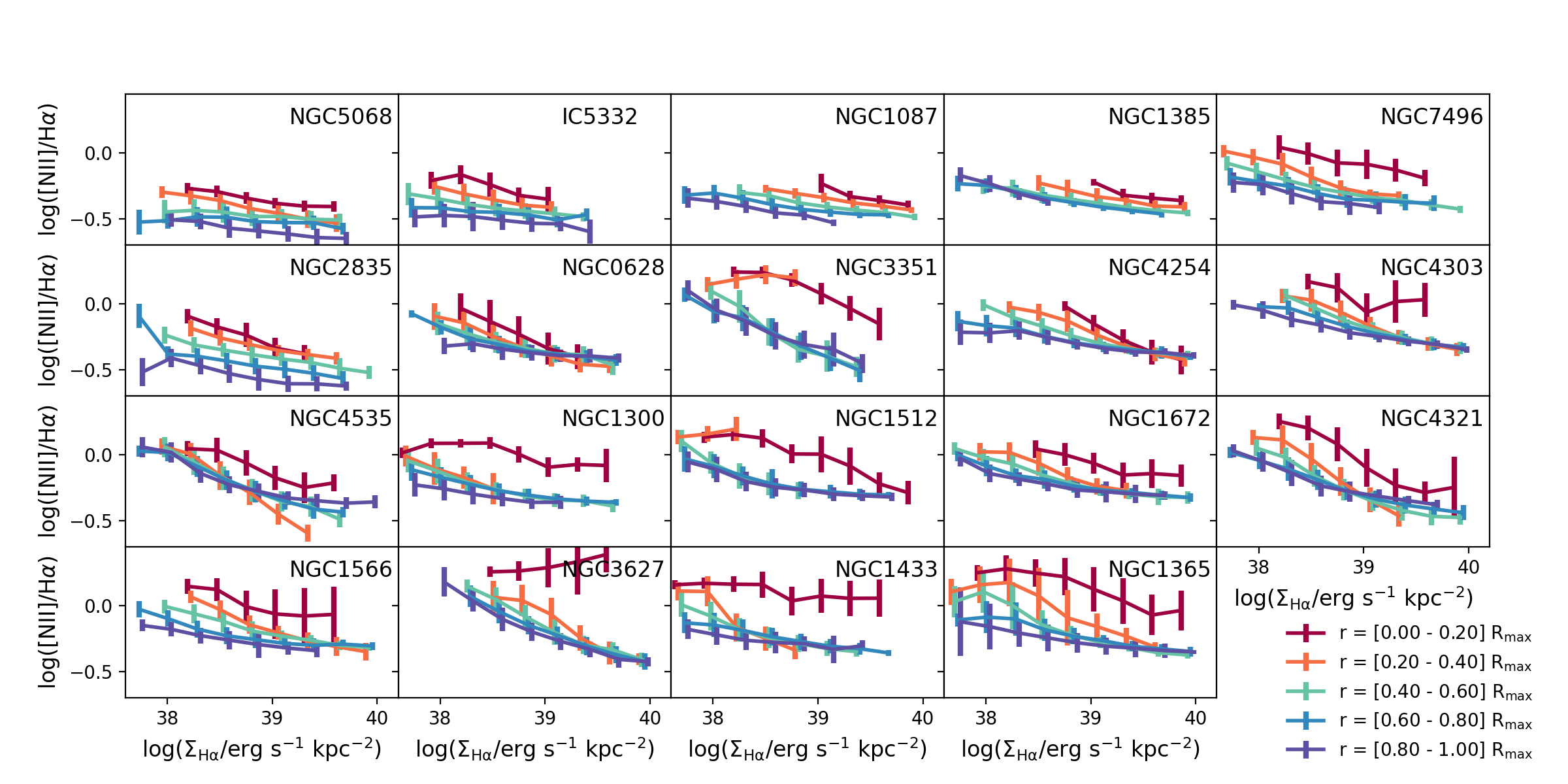}
        \caption{{Same as Fig.~\ref{figS2} but for log(\nii/H$\bm{\alpha}$).}
        Several galaxies (NGC~3351, NGC~1300, NGC~1512, NGC~3627, NGC~1433, and NGC~1365) have exceptionally high values of this (and other) line ratios in their central radial bin (in red). Aside from this feature, the galaxies in the sample follow an average relation between log(\nii/\ha) and \SHa, with a clear secondary dependence on metallicity. Inner regions have higher metallicity and show elevated \nii/\ha\ at fixed \SHa.}
        \label{figN2}
\end{figure*}

\begin{figure*} 
        \includegraphics[width=0.99\textwidth, trim=0 0 0 0, clip]{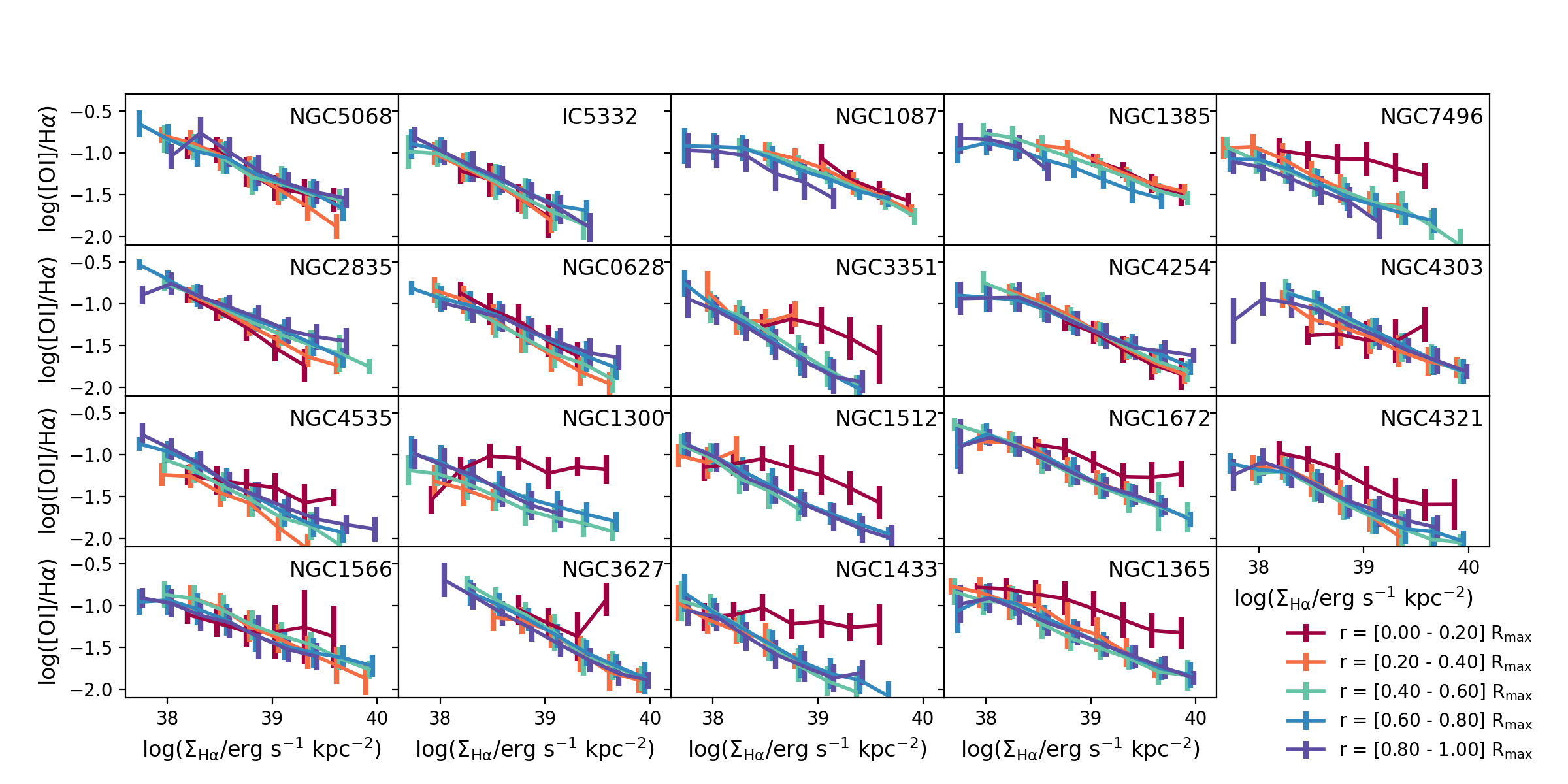}
        \caption{{Same as Fig.~\ref{figS2} but for log(\oi/H$\bm{\alpha}$).} Several galaxies (NGC~3351, NGC~1300, NGC~1512, NGC~3627, NGC~1433, and NGC~1365) show exceptionally high line ratio values in their central regions (in red).}
        \label{figO1}
\end{figure*}

\subsubsection{Model predictions for low-ionisation lines}
\label{sec:low_lines_models}

Before describing the data, it is helpful to consider the expected trends in the scenario where the DIG consists of low-$U_0$ radiation-bounded clouds, subject to the diluted ionising spectrum of young stellar populations. Under these conditions the DIG is effectively described by the same model ingredients as canonical \hii\ regions, albeit with lower $\log{U_0}$. In Fig.~\ref{fig:models_lowion}, we show the predictions for the \sii/\ha, \nii/\ha\ and \oi/\ha\ line ratios from gas photoionised by a 2~Myr SSP generated with the flexible stellar population synthesis (\textsc{fsps}) code. As discussed in \cite{Byler2017}, the line ratios obtained from a 2~Myr old `burst' model are roughly equivalent to those expected from a model with a constant star-formation history, once it has reached a steady state. Moreover, for the adopted MIST isochrones the line ratios do not change substantially for ages between 1 and 4~Myr (see Fig.~24 of \citealt{Byler2017}). 

The photoionisation model calculations are run using the \textsc{cloudy} photoionisation code v17.02, last described by \cite{Ferland2017}, and more details on the input parameters are given in Appendix~\ref{app:photomodels}. For each line ratio we show its dependence on the ionisation parameter log$U_0$, for different values of oxygen abundance, $\mathrm{12+log(O/H)} = [8.09, 9.09]$ in intervals of $0.2$~dex, where the solar oxygen abundance is taken to be $\mathrm{12+log(O/H)} = 8.69$ \citep{Asplund2009}. The absolute scale of ISM metallicity is notoriously difficult to determine accurately from \hii\ region line ratios (see e.g. \citealt{Blanc2015, Kewley2019}). Making use of the \cite{Pilyugin2016} metallicity calibration, which is found to be in good agreement with abundances measured via the auroral line temperatures, the \hii\ regions in PHANGS--MUSE galaxies span the metallicity range $\mathrm{12+log(O/H)} = [8.3, 8.6]$ \citep{Kreckel2019}. Calibrations that directly compare observed line ratios with photoionisation models, however, predict considerably higher abundances (by ${\sim}0.2{-}0.4$~dex), especially at the high-metallicity end. Considering these systematic uncertainties, we computed models for a range in metallicity wide enough to span the entire range of plausible abundances present in our galaxies.

\sii/\ha, \nii/\ha\ and \oi/\ha\ are predicted to generally increase with decreasing log$U_0$ (i.e. moving towards lower \SHa), although the trends can be seen to flatten or mildly invert for the lowest ionisation parameters and highest-metallicity models. For both \sii/\ha\ and \oi/\ha\ the highest values are associated with near-solar metallicity, while higher and lower metallicities correspond to lower line ratios. \nii/\ha, in contrast, increases monotonically with metallicity, as expected given its use as a metallicity diagnostic. In Fig.~\ref{fig:models_lowion} we show as a light grey horizontal band the range of observed line ratios (from the 10$^{\rm th}$ to the 90$^{\rm th}$ percentile of the distribution) in the DIG of our sample. The dark grey box corresponds to the intersection with the expected log$U_0$ in the DIG ($\log{U_0} = [-4.5, -3.5$]).

The absolute values of the line ratios observed in our data also agree, to first order, with those predicted by the models (as seen by comparing the models with the grey boxes in Fig.~\ref{fig:models_lowion}). The match is not perfect, however: our models do not reach the highest values of \nii/\ha\ observed in the DIG at the lowest \SHa\, and in the range of ionisation parameters expected in the DIG, the \nii/\ha\ ratio does not increase, but seems roughly flat. We do not consider these discrepancies large enough to discredit the assumption that the DIG is ionised by leaked radiation from \hii\ regions, although such a discrepancy has been used in the past to motivate the need for an additional shock component to the DIG \citep{Haffner2009}.

The models presented here do not include changes in the N/O abundance ratio, age of the ionising clusters, or hardening of the ionisation spectrum due to intervening absorption, effects that would increase the values of the considered line ratios and potentially lead to a better match with the data. Hardening of the spectrum due to intervening absorption will cause an increase in \nii/\ha\ and \sii/\ha\ with respect to the models shown in Fig.~\ref{fig:models_lowion}. A change in N/O will manifest itself mostly in an increase in the flux of the \nii\ line and a higher \nii/\ha\ for all values of the ionisation parameter. We discuss the effect of adding a radiation source with a harder spectrum in Sect.~\ref{sec:holmes_models}.

\subsubsection{Observed trends for low-ionisation lines in the DIG}
\label{sec:low_ion_data}

In Fig.~\ref{figS2} we show log(\sii/\ha) as a function of \SHa\ for regions included within the DIG mask for all the galaxies in our sample. Each galaxy is sub-divided into five radial bins, going from the galaxy centre (red) to the outskirts (purple) in steps of $0.2\,R_{\rm max}$, where $R_{\rm max}$ is a measure of the maximum radial coverage for each galaxy. On average $R_{\rm max} \sim 0.6\,R_{25}$ or ${\sim}8$~kpc for our sample, meaning each radial bin is ${\sim}1.5$~kpc in width. Each galaxy has a different metallicity gradient \citep{Kreckel2019}, so stepping in relative radius aims to qualitatively capture radial variations in metallicity. The line ratios are not corrected for extinction in the DIG. Such a correction would have a small impact since we only consider ratios of lines close in wavelength.

The relation between \sii/\ha\ and \SHa\ is remarkably similar among different radial bins and across different galaxies. At low \SHa\ (i.e. $\log(\SHa/\mathrm{erg~s^{-1}~kpc^{-2}}) < 38$) we find $\log(\sii/\ha) \sim 0.0$, while at the highest \SHa\ [$\log(\SHa/\mathrm{erg~s^{-1}~kpc^{-2}}) \sim 40$] we find $\log(\sii/\ha) \sim {-}0.5$. Variations in \sii/\ha\ with radius within individual galaxies are small and are most evident at high \SHa, especially for NGC~628 and NGC~4254. These variations go in the  direction of higher-metallicity inner regions having lower \sii/\ha\ at fixed \SHa, which is the trend expected from the photoionisation models, assuming these massive galaxies have super-solar metallicities in their central regions. 

The most evident deviations from the general trend are seen in a set of five galaxies, NGC~3351, NGC~1300, NGC~1512, NGC~3627, and NGC~1433, which show enhanced line ratios at given \SHa\ in their central regions. This behaviour is seen also in \nii/\ha\ and \oi/\ha\ and even more clearly in the high-ionisation \oiii/\hb\ line ratio discussed in the next section. We therefore postpone a discussion of the origin of this trend to Sect.~\ref{sec:holmes_models}.

In Figs.~\ref{figN2} and~\ref{figO1} we show the changes in \nii/\ha\ and \oi/\ha\ with \SHa, presented in the same format as in Fig.~\ref{figS2}. \nii/\ha\ reveals its metallicity dependence through the vertical offset between radial bins, with the innermost radius being highest in \nii/\ha\ (particularly evident in e.g. NGC~2835). Such patterns demonstrate the existence of a metallicity gradient, such that the  metal-rich inner regions have higher \nii/\ha\ in their DIG. In the case of  \oi/\ha\ all radial bins, except for the inner regions of massive galaxies, follow a similar trend and no strong metallicity dependence is observable. This is also in accordance with model predictions, which show that \oi/\ha\ is only marginally affected by metallicity except for $\mathrm{12+log(O/H)} > 8.9$.

\begin{figure*} 
        \includegraphics[width=0.99\textwidth, trim=0 0 0 0, clip]{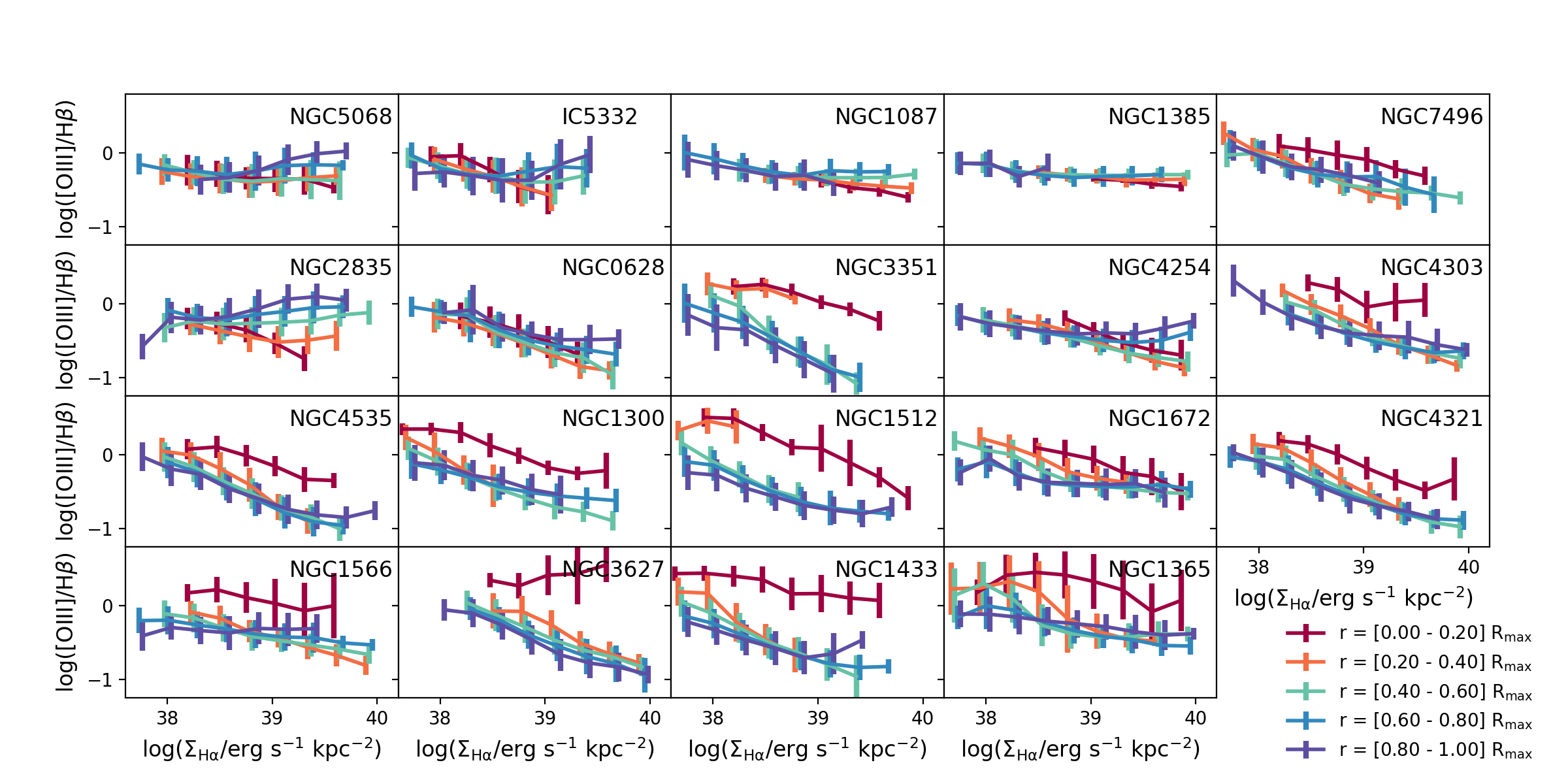}
        \caption{{Log(\oiii/H$\bm{\beta}$) in the DIG as a function of ${\log(\SHa/\mathrm{erg~s^{-1}~kpc^{-2}})}$ for different radial bins for individual galaxies in the PHANGS--MUSE sample.} Galaxies are ordered by stellar mass from lowest (top left) to highest (bottom right). Several galaxies (NGC~3351, NGC~1300, NGC~1512, NGC~3627, NGC~1433, and NGC~1365) show exceptionally high line ratio values in their central regions (in red). Galaxies present a systematic change in the slope of the log(\oiii/\hb) versus \SHa\ relation, with low-mass galaxies showing comparable line ratios at low and high \SHa \ and high-mass galaxies showing a clear decrease in log(\oiii/\hb) with \SHa. Both of these trends are at odds with predictions from photoionisation models.}
        \label{figO3}
\end{figure*}

\subsection{High-ionisation line ratios in the DIG:  \texorpdfstring{\oiii/\hb}{[OIII]/Hbeta}}
\label{sec:high-ion}

Literature studies of the change of \oiii/\hb\ with \SHa\ in the DIG  present conflicting results, with some authors finding little or no change (for the  Milky Way, \citealt{Haffner2009}; for M31, \citealt{Greenawalt1998}) and others finding an increase in the DIG (with scale height in edge-on galaxies, e.g. \citealt{Rand1998, Otte2001, Collins2001}, or with decreasing \SHa\ towards the inner regions of the Milky Way, e.g. \citealt{Madsen2005}). Either trend contradicts models for clouds photoionised by massive stars, since such models predict \oiii/\hb\ to decrease with decreasing log~$U_0$. This discrepancy has led to proposals of additional ionisation sources in the DIG, for example shocks \citep{Collins2001}. An important piece of evidence was uncovered by the analysis of \cite{Zhang2017} based on MaNGA data. They found that the change of \oiii/\hb\ with \SHa\ depends on stellar mass. Low-mass galaxies show a marginal decrease in \oiii/\hb, while massive galaxies show a clear increase in \oiii/\hb\ with decreasing \SHa. 

In Fig.~\ref{figO3} we reassess these trends using the PHANGS--MUSE sample. Following the same formalism as in Figs.~\ref{figS2} to~\ref{figO1}, we show the change of \oiii/\hb\ with \SHa\ for different radial bins in individual galaxies. As previously noted for low-ionisation line ratios in Sect.~\ref{sec:low_ion_data}, the inner regions of several massive galaxies (NGC~3351, NGC~1300, NGC~1512, NGC~3627, NGC~1433, and NGC~1365) have noticeably higher \oiii/\hb\ at fixed \SHa\ than all other data. In addition, we find systematic changes of the observed \oiii/\hb\ versus \SHa\ trends with stellar mass. Lower-mass galaxies show a roughly constant \oiii/\hb\ as a function of \SHa\ (e.g. NGC~5068, IC~5332, NGC~1087, and NGC~1385). For higher-mass galaxies, in contrast, we find a decrease in \oiii/\hb\ as a function of \SHa\ of up to 1~dex from $\log(\SHa/\mathrm{erg~s^{-1}~kpc^{-2}}) = 38{-}40$. These trends agree with the results of \citet{Zhang2017}, obtained using kiloparsec-resolution MaNGA data.


 \subsection{Model predictions for HOLMES}
 \label{sec:holmes_models}

 \begin{figure*} 
     \centering
     \includegraphics[width=0.9\textwidth, trim=0 0 0 0, clip]{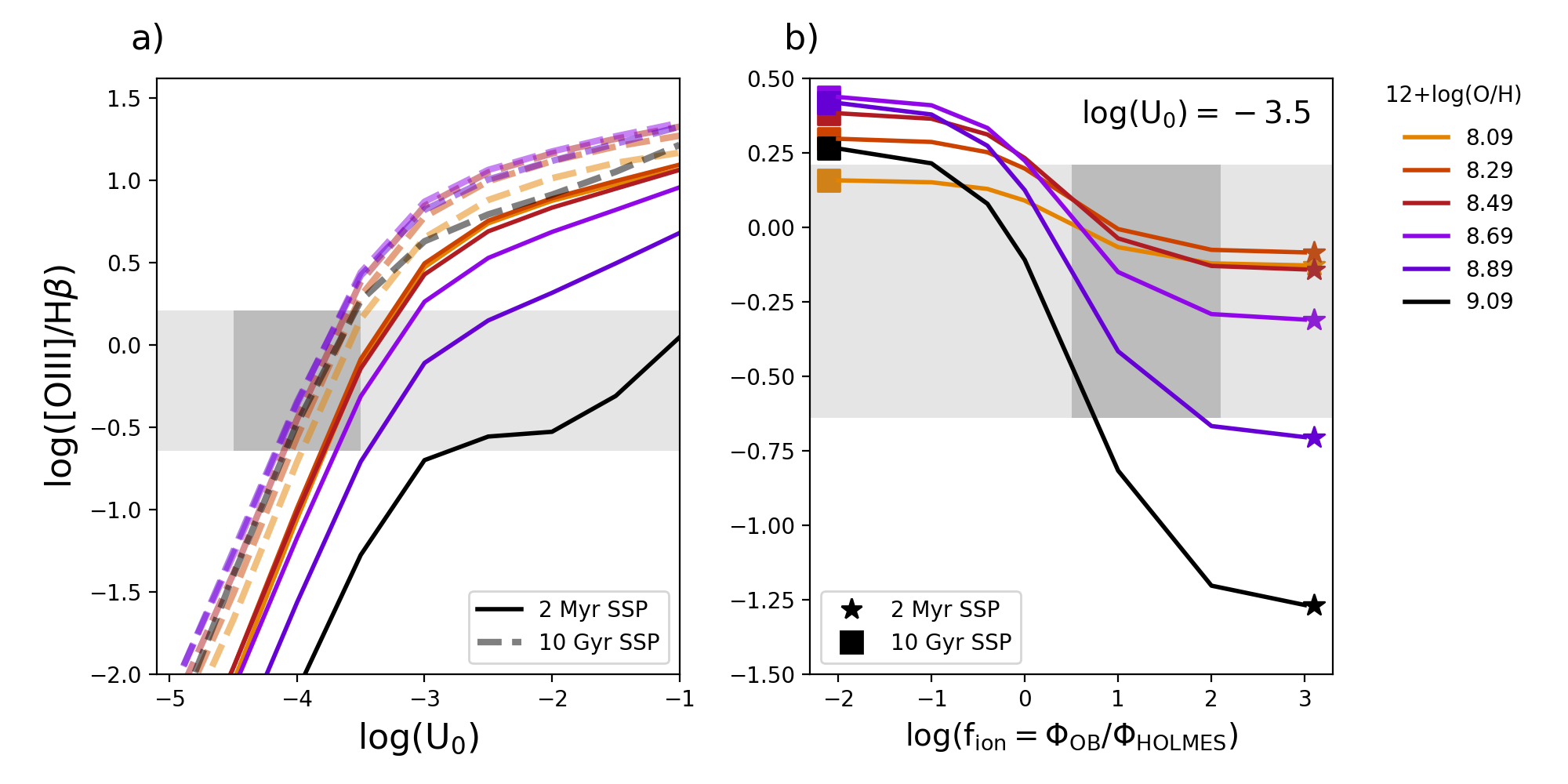}
        \caption{ {Model predictions for the \oiii/\hb\ line ratio considering both young stars and HOLMES.} a) Model predictions for photoionised clouds, computed with \textsc{cloudy} and a 2~Myr (solid lines) and a 10~Gyr (dashed lines) SSP spectrum. The line ratio log(\oiii/\hb) is plotted as a function of the ionisation parameter at the incident face of the cloud ($\log(U_0)$). The vertical extent of the grey area highlights the range of line ratios observed in the DIG in the PHANGS--MUSE data. Different colours correspond to different gas-phase metallicities, with the same colour-coding in both panels, as indicated in the legend. All models predict log(\oiii/\hb) to increase with \SHa, in contrast to what is observed.
        b) Model predictions for the log(\oiii/H$\bm{\beta}$) line ratio as a function of the fraction of the ionising radiation originating from old stars, ${f_\mathrm{ion} = \Phi_\mathrm{OB} / \Phi_\mathrm{HOLMES}}$.We show models for a fixed value of $\log(U_0)= -3.5$ and different metallicities. The horizontal light grey band corresponds to the range of values of log(\oiii/\hb) observed in the DIG of our sample, while the dark grey box corresponds to the range of $f_\mathrm{ion}$ observed in the PHANGS data. $f_\mathrm{ion}$ is estimated by computing the ionising photons expected to be produced by HOLMES given the local stellar mass surface density.}
        \label{fig:models_highion}
 \end{figure*}

 In the left panel of Fig.~\ref{fig:models_highion} we show the predicted changes in \oiii/\hb\ for clouds photoionised by a 2~Myr SSP (solid lines) as a function of $\log U_0$ (same model parameters as Fig.~\ref{fig:models_lowion}). \oiii/\hb\ is predicted to increase with $\log U_0$, at odds with our observations where this ratio remains constant or decreases with \SHa. The light grey horizontal band corresponds to the range between the 10$^{\rm th}$ and the 90$^{\rm th}$ percentile of the observed \oiii/\hb\ in the DIG, and only overlaps with the models at $\log U_0 \sim {-}3.5$. The models, however, do not cover the data for the range of expected values of $\log U_0$ in the DIG (dark grey box). In fact, the only similarity between models and data can be found at the high \SHa\ end, where more metal-rich, higher-mass galaxies show lower \oiii/\hb, as predicted by the models.  An additional ionisation source, which would contribute little to the Balmer line emission, but generate a higher \oiii/\hb\ ratio, is therefore needed to explain the data. 

 Several authors have pointed out that there is a natural candidate for a pervasive ionisation source with the necessary hard spectrum: HOLMES \citep{Binette1994, Stasinska2008,Flores-Fajardo2011, Sarzi2010, Yan2012, Belfiore2016a, Byler2019}.
\cite{Byler2019} presented photoionisation models for an old stellar population based on MIST stellar evolution tracks, which provide a consistent framework for the evolution of low- and intermediate-mass stars from the main sequence to the white dwarf cooling sequence and are therefore very well-suited to the study of the ionising spectrum of HOLMES. We build on the work of \cite{Byler2019} and generate a 10~Gyr spectrum using \textsc{fsps} with MIST isochrones. This spectrum is used to generate a grid of \textsc{cloudy} models, covering the same values of gas phase metallicity and ionisation parameter as our 2~Myr models. A full description of the models is presented in Appendix~\ref{app:photomodels}. 

The \oiii/\hb\ sequence  produced by the 10~Gyr SSP models are shown as dashed lines in Fig.~\ref{fig:models_highion}a. At solar metallicity and above, the increase in \oiii/\hb\ with respect to the 2~Myr models is ${>}0.5$~dex, and gets as large as $1.5$~dex at $\mathrm{12+log(O/H)} = 9.09$. The old star models are able to produce sufficiently high \oiii/\hb\ at low $\log U_0$, as evidenced by the fact that their position now overlaps with the dark grey box (expected position of the data) in Fig.~\ref{fig:models_highion}a. Since the \oiii/\hb\ ratios predicted by young star models are very low, even a small fraction (a few percent) of ionising flux from HOLMES can dominate the \oiii\ emission at low \SHa.

\subsection{DIG as a mixing sequence: From leaking radiation to HOLMES}
\label{sec:mixing_models}

We next explore the possibility that the DIG is ionised by a combination of ionising photons from both young stars (which we represent with the 2~Myr SSP model) and HOLMES (represented by the 10~Gyr model). We build \textsc{cloudy} models with variable ratios of the ionising flux from OB stars with respect to that from HOLMES. We parametrise the models via 
$ f_\mathrm{ion} = \Phi_\mathrm{OB}/ \Phi_\mathrm{HOLMES}$, where we take $\Phi_\mathrm{OB}$ and $\Phi_\mathrm{HOLMES}$ are the ionising fluxes from the 2~Myr and 10~Gyr models, respectively. We compute models with $\log(f_\mathrm{ion}) = [-3, -2, -1, 0, 1, 2]$.
In order to demonstrate the effect of this mixing sequence on \oiii/\hb, we pick models with $\log{U_0} = -3.5$ (which we assume typical of the DIG, as discussed in Sect.~\ref{sec:lines_intro}) and present the change of the line ratio as a function of $f_\mathrm{ion}$ for different metallicities in Fig.~\ref{fig:models_highion}b.
 
For $\log(f_\mathrm{ion}) > 2$, one recovers the line ratios predicted by the pure 2~Myr models. A contribution of HOLMES larger than 1\%, however, leads to noticeable changes in the line ratios. Namely, for $-1 < \log(f_\mathrm{ion}) < 2$, the predicted line ratios are intermediate between those of the 2~Myr and 10~Gyr populations, while for lower and higher values they are indistinguishable from those generated by the two extremes. \oiii/\hb\ increases with an increasing contribution of HOLMES to the ionising flux, and the changes in the line ratio are largest for the high-metallicity models. At $\mathrm{12+log(O/H)} = 9.09$, the difference in \oiii/\hb\ between the OB stars and HOLMES models reaches $1.5$~dex. However, at sub-solar metallicity the change in \oiii/\hb\ between the 2~Myr and 10~Gyr models is much more modest (${\sim}0.3$~dex for $\mathrm{12+log(O/H)} = 8.09$). 

In order to compare the data with these model predictions, we estimated $f_\mathrm{ion}$ by computing $\Phi_\mathrm{HOLMES}$ in each spaxel of the MUSE maps from the stellar mass surface density and our fiducial value of the ionising flux of HOLMES (Sect.~\ref{sec:HOLMES_HA}). The $\Phi_\mathrm{OB}$ was then derived from the observed \SHa\ after subtracting the contribution from HOLMES. According to this estimate, the DIG shows a median $\log(f_\mathrm{ion})=1.2$, and the 10$\rm^{th}$ and 90$\rm^{th}$ percentiles of its distribution across the surveyed area are $\log(f_\mathrm{ion}) = 0.5{-}2.0$. This range is shown as the dark grey area in Fig.~\ref{fig:models_highion}b, and the intersection with the observed range of \oiii/\hb, shown as a horizontal light grey band, corresponds quite closely to the location of the models. Remarkably, the  range of observed \oiii/\hb\ can be reproduced assuming a contribution from HOLMES consistent with the expectations from population synthesis modelling of old stellar populations. We therefore confirm that HOLMES are a `natural' solution to explain the DIG line ratios in star-forming galaxies.
 

\subsection{Mixing sequence explains the locus of the DIG in the BPT diagram}
\label{sec:DIG_BPT}

\begin{figure*}[!h] 
    \centering
    \includegraphics[width=0.79\textwidth, trim=0 0 0 0, clip]{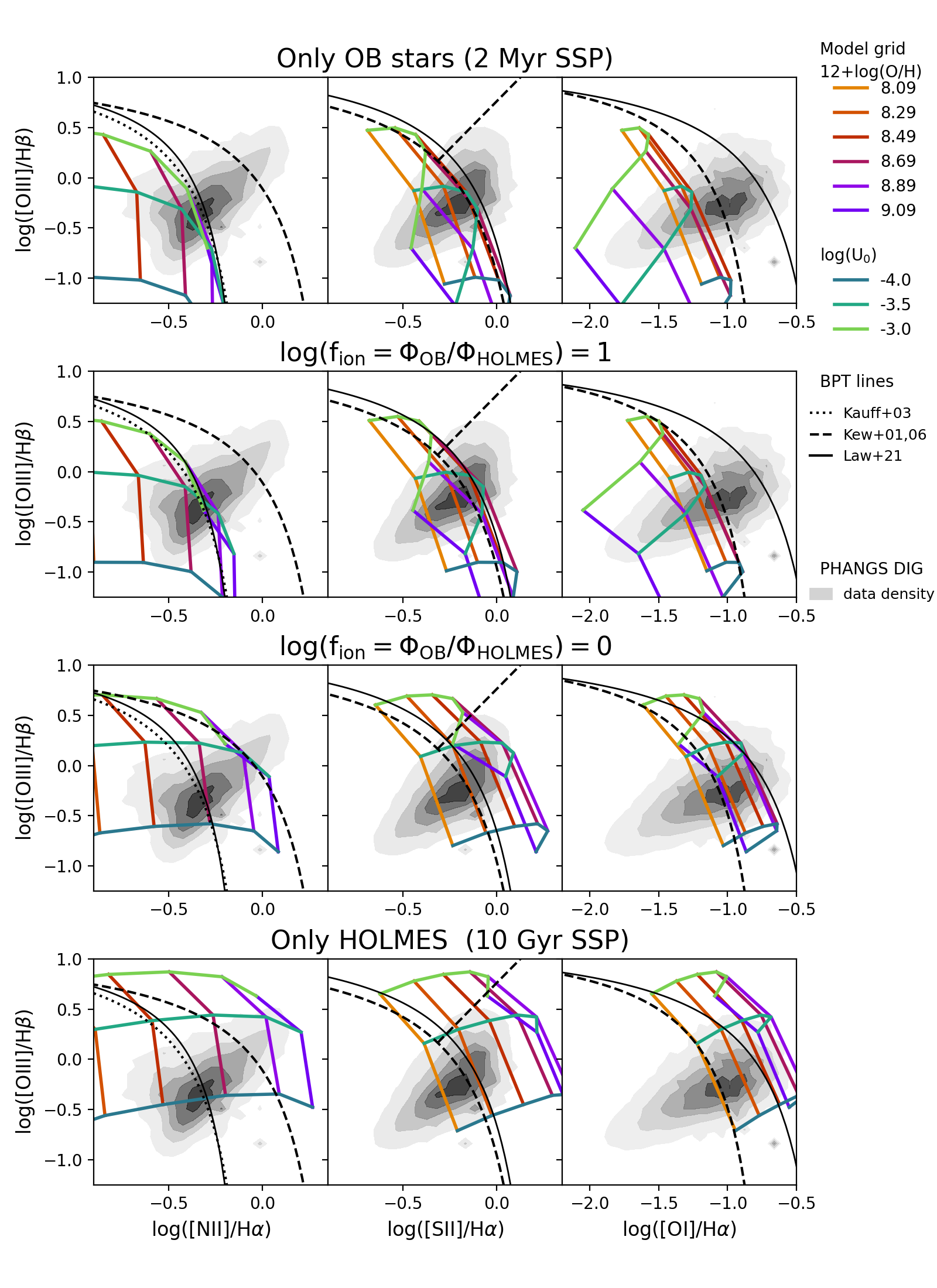}
    \caption{{Position in the \nii\ BPT (left), \sii\ BPT (middle), and \oi\ BPT (right) of photoionisation models with varying ${f_\mathrm{ion} = \Phi_\mathrm{OB} / \Phi_\mathrm{HOLMES}}$.} We show models for different metallicity and $\log(U_0) = [-4.0, -3.5, -3.0]$, with colours given in the legend. The top row shows models where the input spectrum includes only OB stars (2~Myr SSP), while in the bottom row the input spectrum only includes old stars (10~Gyr model). The two middle rows illustrate the effect of a decreasing contribution from OB stars to the line ratios in the BPT diagram. Demarcation lines from the literature are shown as dashed \citep{Kewley2001, Kewley2006}, dotted \citep{Kauffmann2003a}, and solid black lines \citep{Law2020}. The grey contours represent the locus of the DIG regions from the galaxies in the PHANGS--MUSE sample (the lowest contour encloses 10\% of the data, and successive contours are in steps of 20\%). The comparison shows that adding a varying contribution of HOLMES allows all the line ratios observed in the DIG to be broadly covered.
    }
    \label{fig:BPT_models}
 \end{figure*}

In Fig.~\ref{fig:BPT_models} we show the position of various models along the mixing sequence in the classical BPT diagrams: log(\oiii/\hb) versus log(\nii/\ha) (\nii\ BPT, left), log(\oiii/\hb) versus log(\sii/\ha) (\sii\ BPT, middle), and log(\oiii/\hb) versus log(\oi/\ha) (\oi\ BPT, right). We present the 2~Myr (top row) and 10~Gyr (bottom row) models, in addition to models along the mixing sequence with $\log(f_\mathrm{ion}) = 1$ and $\log(f_\mathrm{ion}) = 0$. We limit the grid to the values of $\log U_0 = [-4.0, -3.5, -3.0]$, because these models are the ones that more closely match the observed DIG line ratios. Demarcation lines from the literature are shown as dashed \citep{Kewley2001, Kewley2006}, dotted \citep{Kauffmann2003a} and solid black lines \citep{Law2020}. The latter is a recent empirical determination of the BPT demarcation lines based on the MaNGA survey.

In the \nii\ BPT, the original line suggested by \cite{Kewley2001} was not a good representation of the data. Even though a new demarcation line was suggested by \cite{Kauffmann2003a} to provide a closer representation of the sequence of star-forming galaxies in the legacy Sloan Digital Sky Survey (SDSS-I) data, researchers have continued to use the \cite{Kewley2001} demarcation line in the \nii\ BPT diagram because it provides a useful `intermediate' zone between the star-forming sequence and the AGN locus. The empirical \cite{Law2020} line, derived from MaNGA data, matches the \cite{Kauffmann2003a} line very well.

We show the positions of all DIG regions in the three BPT diagrams in Fig.~\ref{fig:BPT_models} as grey contours (the lowest contour encloses 10\% of the data, and successive contours are shown in steps of 20\%). For this comparison we exclude data from NGC~1365 to avoid its prominent AGN ionisation cone. 

Figure~\ref{fig:BPT_models} shows that with decreasing $f_{\rm ion}$ (i.e. larger contribution from HOLMES), models move `diagonally up', towards the top-right corner in all diagrams. $\log(f_{\rm ion}) = 1$ (HOLMES contribute 1/10 of the ionising flux), a small number of models cross the relevant demarcation lines for the star-forming sequence into the area of the BPT diagram associated with LI(N)ERs and AGN. The shift into the LI(N)ER region is almost complete for $\log(f_{\rm ion}) = 0$ (equal contributions of HOLMES and OB stars to the ionising flux). Coincidentally, models with $\log(f_{\rm ion}) = 0$ in the \nii\ BPT provide an excellent match to the old \cite{Kewley2001} line, highlighting the usefulness of this line as a probe for the DIG mixing sequence. 

While the models extend into the top left area of the BPT diagrams, generally associated with Seyfert AGN, that area is occupied by models with $\log{U_{0}} \geq -3$. Qualitatively, the best match between models and data is achieved with $\log{U_0} =[-4.0, -3.5]$ and requires the full range of the mixing sequence between OB stars and HOLMES to explain the line ratios in the top-most right corner of the diagram, especially in the \nii\ and \oi\ BPT diagrams, while models only containing HOLMES at high metallicity over-predict \sii/\ha. Remarkably, considering all the models along the mixing sequence, we can span the entire range of line ratios found in the data.

\begin{figure*} 
        \includegraphics[width=0.9\textwidth, trim=0 0 0 0, clip]{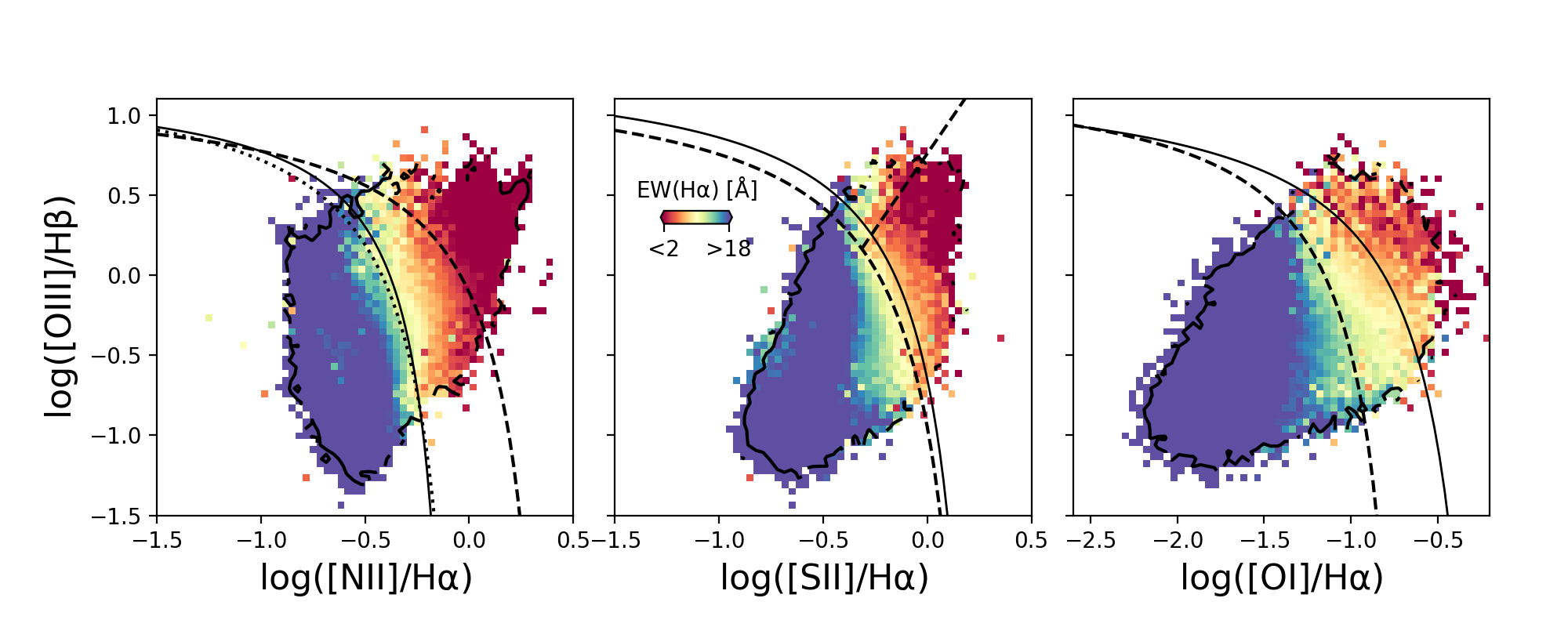}
        \centering
        \caption{{Position of the DIG regions in our sample in the \nii\ BPT (left), \sii\ BPT (middle), and \oi\ BPT (right) colour-coded by the median EW(H$\bm{\alpha}$).} We excluded NGC~1365 because of its prominent AGN cone. Demarcation lines from the literature are shown as dashed \citep{Kewley2001, Kewley2006}, dotted \citep{Kauffmann2003a}, and solid black lines \citep{Law2020}. The colour bar for EW(\ha) highlights the transition region between the 16$^{\rm th}$ and 84$^{\rm th}$ percentiles of the EW(\ha) distribution of all regions.}
        \label{fig:BPT_ewha}
\end{figure*}

In Fig.~\ref{fig:BPT_ewha} we show the location of the DIG regions for the entire sample (again excluding NGC~1365) in the three BPT diagrams colour-coded by the median EW(\ha) in each bin, as a proxy for the relative importance of HOLMES. As a reminder, we expect $\mathrm{EW}(\ha) \sim 1.1$~\AA\ for pure HOLMES under our fiducial assumptions. We observe a strong correlation between EW(\ha) and the position of DIG regions in the three BPT diagrams, with EW(\ha) decreasing diagonally up in excellent agreement with the expectations of an increased contribution of HOLMES moving from the star-forming locus to the LI(N)ER regime. \cite{Lacerda2018} and \cite{Law2020} also showed similar trends between EW(\ha) and the position of regions in the BPT diagram, but this work demonstrates unambiguously that this trend is also present within the DIG in the disc of star-forming galaxies.

 \section{Discussion}
 \label{sec:discussion}

 \subsection{Mean free path for ionising radiation}
\label{Sec:mfp}
A key challenge for models that consider leaking radiation as the ionising source for the DIG is to explain the propagation of ionising photons within the multi-phase ISM. Early Monte Carlo 3D photoionisation models were based on the assumption of a uniform medium and needed to assume a lower midplane gas density than observed in order to fully ionise the extra-planar DIG \citep{Wood2004}.
 More recently, works coupling 3D photoionisation with more realistic ISM structures (e.g. from hydrodynamical simulations) have demonstrated that low-density chimneys naturally emerge, and allow relatively unimpeded transport of ionising photons away from the midplane \citep{Wood2010, Barnes2015, Kado-Fong2020}. 

In particular, a mean free path for ionising photons of the order of $\sim$1 kpc implies a small filling factor of neutral gas in the galaxy disc. For $13.6$~eV photons, an optical depth $\tau \sim 1$ is reached after a neutral column density of ${\sim}1.6 \times 10^{17}~{\rm cm}^{-2}$. For our best-fit value for the mean free path of $1.9$~kpc, we obtain a mean atomic hydrogen number density $n_{\rm H}$ of ${\sim} 3 \times 10^{-5}~{\rm cm}^{-3}$. More energetic photons are less affected, but even for 30~eV photons, the estimated mean free path implies $n_{\rm H} < 10^{-3}~{\rm cm}^{-3}$ on average. Our model, therefore, requires most of the volume of the disc to be filled with either very ionised or very diffuse gas. This is in qualitative agreement with the fact that most of the DIG emission does not originate from the region where the cold gas disc lies (scale height of ${\sim}100$~pc; \citealt{Ferriere2001, Kalberla2009, Heyer2015}), but from a thicker ionised gas layer, where the filling factor of neutral material is expected to be significantly lower. It has also been suggested that  young regions the Milky Way could be situated above or below the disc midplane \citep{Alves2020}, meaning that photons escaping such regions may be less hampered by neutral gas.

 \subsection{Spectral hardening of leaking radiation and alternative heating sources}
 
\begin{figure*} 
    \includegraphics[width=0.97\textwidth, trim=0 0 0 0, clip]{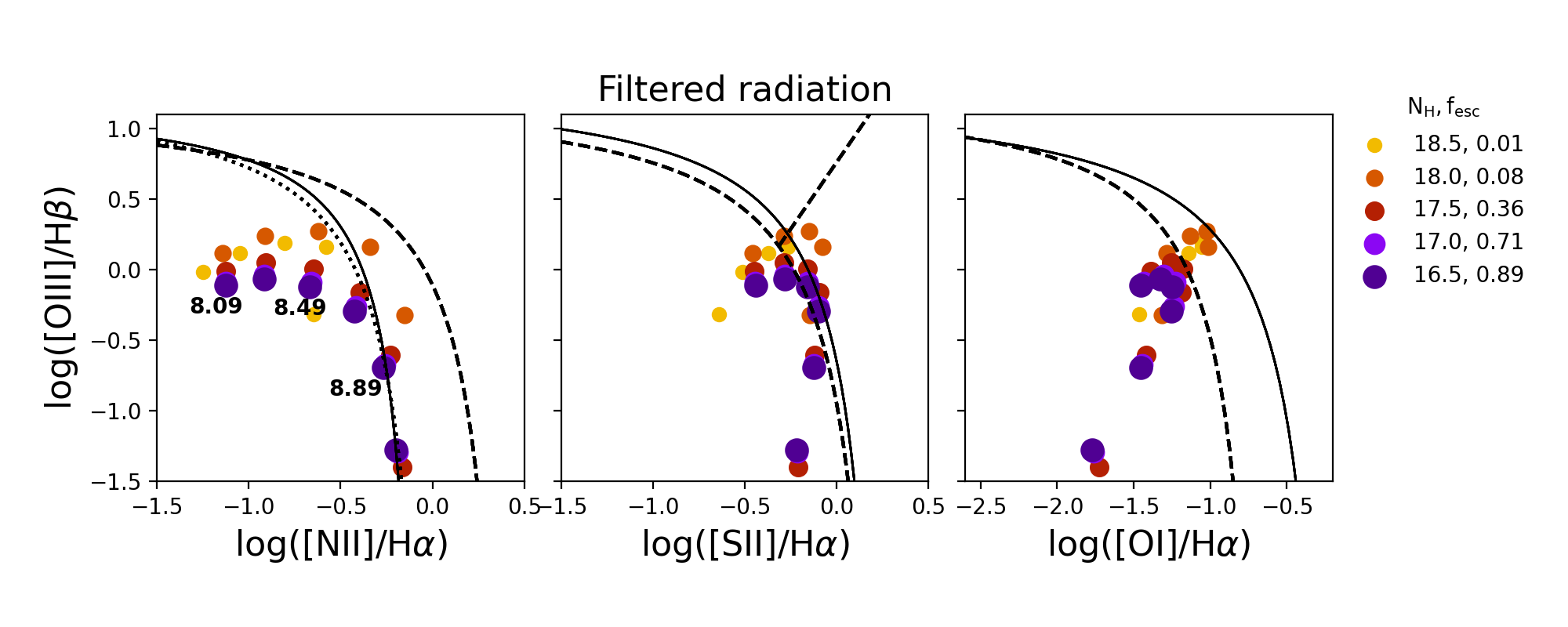}
    \centering
    \caption{{Photoionisation models of clouds illuminated by a hardened radiation field leaking from \hii\ regions in the \nii\ BPT (left), \sii\ BPT (middle), and \oi\ BPT (right) diagrams.} The input spectrum is a 2~Myr SSP, which has been transmitted from a density-bounded \hii\ region. We consider \hii\ regions with a range of hydrogen column densities ($N_{\rm H}$, reported in the legend). The legend also shows the escape fraction from the \hii\ region corresponding to each column density value.  All models shown have $\log{U_0} = -3.5$ and varying metallicity (three representative metallicities are labelled in the left panel to demonstrate that metallicity generally increases along the $x$ axis). Demarcation lines from the literature are shown as dashed \citep{Kewley2001, Kewley2006}, dotted \citep{Kauffmann2003a}, and solid black lines \citep{Law2020}.}
    \label{fig:BPT_filtered}
\end{figure*}

 Considering an inhomogeneous \hii\ region, one can assume two different scenarios for the leakage of ionising radiation. Either the ionised gas shell around a massive cluster presents fully transparent lines of sight, allowing a fraction of the ionising radiation to leak into the ISM unprocessed (the `picket fence' scenario), or radiation leaks if the ionised cloud is density-bounded, that is, if the cloud does not terminate at an ionisation front with the neutral medium but lacks the outer zone. In this case, the ionising spectrum propagating into the ISM will have been partially absorbed by neutral hydrogen and helium in the nebula (the `filtered radiation' scenario). Since the absorption cross-section of hydrogen decreases with frequency,~$\nu$, roughly as $\nu^{-3}$, the escaping radiation will be harder (at least in the energy interval between the ionisation potentials of hydrogen and helium). Real \hii\ regions may lie somewhere in between these limiting cases. The models we have discussed in this work correspond to the picket fence scenario, while in the filtered radiation scenario the spectrum ionising the DIG will be harder than the unprocessed 2~Myr SSP we have used. A harder spectrum will lead to stronger emission from low-ionisation lines but also higher \oiii/\hb\ \citep{Sokolowski1991, Hoopes2003, Flores-Fajardo2011, Zhang2017, Weber2019, Pellegrini2020a}. 

To demonstrate the effect of hardening of the radiation field, we computed a set of density-bounded photoionisation models with the 2~Myr SSP as input, and terminated the models at different values of hydrogen column density, resulting in different escape fractions of ionising radiation. The filtered (transmitted) radiation escaping from these density-bounded regions was then used as input for a second set of \textsc{cloudy} models, aiming to reproduce DIG clouds illuminated by a filtered radiation field. In Fig.~\ref{fig:BPT_filtered}, we show a representative set of such models with $\log{U_0} = -3.5$, and varying metallicity and escape fraction (as shown in the legend). 
 
The dependence of line ratios on escape fraction is non-monotonic. While in general the largest enhancements for both low (e.g. \sii/\ha) and high (\oiii/\hb) ionisation lines occur for the smallest escape fractions,  fluorescent production\footnote{For very low escape fractions the filtered spectrum will have so few ionising photons that, in order to obtain the required ionisation parameter, one requires a large unattenuated flux. The filtered model spectrum will therefore have a large number of photons with enough energy to excite an electron in the hydrogen atom from ground level ($n = 1$) to $n = 3$ or~4 level (fluorescent production).} of hydrogen Balmer lines leads to a decrease in these line ratios for very small (${<}1$\%) escape fractions \citep{Flores-Fajardo2011, Zhang2017}. Our results confirm the conclusions of \cite{Zhang2017}: enhancements in line ratios in hardened radiation models are modest (${<}0.2$~dex) and do not allow leaky radiation models to expand sufficiently far into the LI(N)ER region of the BPT diagram. Moreover, the largest enhancements are found for escape fractions of ${\sim}10\%$, which are not sufficient to power the DIG.

In order to compute the more complex process of radiation hardening in the DIG itself as a function of distance from \hii\ regions one needs to take into account the 3D radiative transfer problem in a clumpy medium, as considered by, for example, \cite{Pellegrini2020}. In this framework, the DIG is assumed to be ionised by (filtered) radiation from \hii\ regions, and the clumpy ISM will allow propagation of ionising photons along low-density paths. In this case, one may think of the whole galactic disc as a `Str{\"o}mgren volume', and the line ratios changing within this volume as they would within an \hii\ region. In analogy to the structure of \hii\ regions, low-ionisation line ratios (\sii/\ha, \nii/\ha, and \oi/\ha)  increase towards the edge of the Str{\"o}mgren volume, while \oiii/\hb\ decreases. \cite{Weber2019} computed 3D photoionisation models in a clumpy medium and quantitatively demonstrate this point. This effect, however, cannot explain the increase in \oiii/\hb\ as a function of decreasing \SHa\ that we observe in the DIG.

In summary, hardening of leaking radiation is not sufficient to explain the observed properties of the DIG in our galaxy sample. Predicted line ratios from 1D models do not extend sufficiently into the LI(N)ER regime to explain the more extreme observed line ratios. Moreover, 3D models predict a decrease in \oiii/\hb\ with distance from the ionising sources, therefore, not matching the observed decreasing or flat trend of \oiii/\hb\ versus \SHa.

Additional heating processes and/or alternative sources for a hard spectral component may be able to explain the observations presented in this paper. For example, shocks models \citep{Allen2008, Rich2011} largely overlap with the LI(N)ER section of the BPT diagram. However, we find a trend of the BPT line ratios with EW(\ha) across the entire sample - meaning that the extra heating mechanism needs to be associated with stellar mass density, supporting HOLMES. 

The presence of LI(N)ER emission in barred galaxies deserves a more specific discussion. As a result of their underlying orbital structure, bars are expected to drive gas flows and shocks in the gaseous component. Hydrodynamical models predict the existence of shocks along the leading edge of the bar, encompassing the thin gas (or dust) lanes often observed in barred galaxies \citep{Athanassoula1992, Sormani2015}. The gas within the bar is not expected to be shocked, but often shows enhanced velocity dispersions \citep{Sun2020b} due to high shear \citep{Emsellem2015a}. Bars are also predicted to sweep the regions of the disc inside of the bar radius (the `interbar' region) clear of cold gas, as is indeed observed by inspecting the molecular gas distribution of PHANGS galaxies  \citep{Leroy2021a}. Absence of cold gas and star formation within the interbar region leads to the appearance of so-called star-formation deserts \citep{James2009p, Neumann2020}. Within the PHANGS-MUSE sample, 12 of our targets are classified as barred galaxies by \cite{Querejeta2021} based on \textit{Spitzer} imaging (NGC~1300, NGC~1365, NGC~1433, NGC~1512, NGC~1566, NGC~1672, NGC~3351, NGC~3627, NGC~4303, NGC~4321, NGC~4535, and NGC~7496). Of these, at least six show an extensive star-formation desert in the interbar region, showing low EW(\ha) and LI(N)ER classification (NGC~1300, NGC~1433, NGC~1512, NGC~1566, NGC~3351, and NGC~3627). While somewhat less conspicuously, the interbar region is associated with lower EW(\ha) and higher fraction of LI(N)ER pixels also in the other barred galaxies. 
        
Overall, the bar exerts an important influence on the distribution of star-formation and line emission in both the bar and interbar region, but HOLMES, rather than shocks, seem to be the likely cause of the observed line ratios in the interbar region.
Moreover, our proposed solution relying on HOLMES has the advantage of not requiring fine-tuning, since the observed line ratios are reproduced by adding the predicted amount of ionising photons from HOLMES.

\subsection{Consequences for SFR measurements}
\label{sec:SFR}

We have demonstrated that the DIG is powered primarily by leaking radiation from \hii\ regions, and only in small part by a spatially diffuse, harder spectral component originating from HOLMES. This conclusion argues in favour of including the \ha\ contribution from the DIG in the total computation of the integrated star-formation rate (SFR) in galaxies. As a consequence, the inability to spatially map the leaking photons from \hii\ regions does not prevent accurate SFR measurements from observations at kiloparsec resolution (as in e.g. MaNGA). However, the significant escape fraction from \hii\ regions does caution against applying standard SFR calibrations, which assume no photon loss, to individual \hii\ regions. 

If we calculate the expected contribution from HOLMES to the \ha\ flux directly from the stellar mass surface density, the average contribution of HOLMES to the total \ha\ flux within area mapped by our MUSE data is 2\% (0.9\% if one uses the ionising flux predicted by \textsc{fsps} for a 10~Gyr SSP). NGC~1433 shows the highest fractional contribution of HOLMES to the total \ha\ emission in our sample, with a value of 9\%. The \ha\ flux here is taken to include both \hii\ regions and DIG, but not correcting for extinction. For comparison, the fraction of \ha\ flux in the LI(N)ER region of the BPT diagram is on average 4\%. 
These numbers may be interpreted remembering that our analysis of line ratios in the BPT diagram demonstrates that a contribution from HOLMES of ${>}10$\% to the total number of ionising photons can push the line ratios in the DIG into the LI(N)ER region of the BPT diagram, as shown in Fig. \ref{fig:BPT_models} . Spaxels falling the in the LI(N)ER section of the BPT diagram therefore populate the mixing sequence between ionisation from OB stars and HOLMES, and may have contributions from both ionising sources. Taking the numbers derived above at face value, we conclude that in our sample of galaxies about half of the \ha\ emission in the LI(N)ER part of the BPT diagram is powered by ionising photons originating from OB stars.

The contribution of HOLMES to the \ha\ flux on integrated scales is small in our sample, and is therefore not likely to represent a significant correction to the SFR estimation of nearby galaxies in the blue cloud. To confirm this, we consider a galaxy on the star-formation main sequence as parametrised by \cite{Leroy2019}. In this parametrisation the main sequence for local galaxies has a sub-linear slope of 0.68, leading to an estimated larger contribution of HOLMES at the high stellar masses with respect to the low-mass end. The predicted contribution from HOLMES is nonetheless very small, 0.5\% for galaxies of $\log(M_\star/M_\odot) = 9$ going up to 2\% for galaxies with $\log(M_\star/M_\odot) = 11$. As galaxies move below the main sequence, the contribution of HOLMES will become more important \citep{Belfiore2018}, and eventually become the dominant, and potentially only source of \ha\ line emission on the red sequence. We conclude that the SFR measured from \ha\ by kiloparsec-resolution surveys (e.g. CALIFA, MaNGA, SAMI) are likely to be robust to the small correction due to HOLMES, especially if they exclude low EW(\ha) regions from the computation, and, thanks to their low resolution, already include the contribution from radiation leaking from \hii\ regions.

\subsection{Line ratios and DIG correction}

The fact that line ratios in the DIG change systematically with \SHa\ was already evident in both Galactic and extra-galactic work \citep{Madsen2006, Zhang2017}. Here we confirmed and discussed these trends for galaxies of different masses (and therefore metallicities) and at a spatial resolution sufficient to  distinguish DIG and \hii\ regions. These trends are of particular significance in the context of determinations of the chemical abundances of \hii\ regions. Given the lack of a comprehensive modelling framework for the DIG, several authors have attempted to empirically correct the line ratios of \hii\ regions for DIG contamination \citep{Sanders2017, ValeAsari2019, Espinosa-Ponce2020}. Despite a large body of detailed work on individual galaxies (e.g. NGC~891; \citealt{Rand1998}), IFS for a representative sample of galaxies and their DIG component is currently limited to the kiloparsec-resolution surveys (e.g. CALIFA, SAMI, MaNGA). Such studies can only access uncontaminated DIG line ratios at large distances from \hii\ regions, corresponding to the lowest \SHa\ levels probed in this work. 
  
\cite{ValeAsari2019}, for example, define the DIG in terms of EW(\ha) ($\mathrm{EW}(\ha) < 6$~\AA) using kiloparsec-resolution MaNGA data and argue that the inclusion of the DIG contamination makes little to no difference to the derived abundances. This conclusion is a direct consequence of the fact that the fraction of \ha\ line emission due to HOLMES is small in typical main sequence galaxies (Sect.~\ref{sec:SFR}), but it does not address the impact of the change in lines ratios in dominant component of the DIG due to leaking radiation. The impact of leaking radiation on abundance measurement on kiloparsec scales is likely to be subtle. Focusing on the DIG component due to leaking radiation, this gas is likely to have lower $\log{U_0}$ than \hii\ regions. However, metallicity calibrations implicitly assume a relation between metallicity and ionisation parameter for \hii\ regions, which will not apply to the DIG. Moreover, tools such as \textsc{izi} \citep{Blanc2015, Mingozzi2020}, which directly compare model grids with data, do not allow for the possibility of mixing models with different ionisation parameters. The corrections to integrated metallicities presented in \cite{ValeAsari2019}, which only take into account the negligible fraction of the DIG powered by HOLMES, are therefore likely in need of revision.
  
\cite{Sanders2017}, in contrast, assume the line ratios observed in the lowest \SHa\ regions of MaNGA galaxies to be representative of the DIG component in its entirety (i.e. at all \SHa\ levels). They further assume that the DIG contributes a substantial fraction of the total \ha\ emission (on average 55\%), based on the SINGS \ha\ narrow band survey of nearby galaxies \citep{Oey2007}. While the latter assumption is in agreement with our data for DIG powered by leaking radiation, the former one associates its line emission with extreme line ratios, generally found only in the part of the DIG ionised by HOLMES. These are easy to identify observationally, since they occupy a large area on sky, but contribute a negligible fraction of the \ha\ flux. Using these assumptions, \cite{Sanders2017} compute corrections for the mass--metallicity relation derived from integrated spectra, which can be as large as $0.3$~dex, but could potentially be overestimated.

Based on the results presented here, the PHANGS dataset will allow the validity of these DIG corrections to be tested by taking the fraction of the \ha\ flux and the variations in the line ratios as a function of \SHa\  into account.

\section{Summary and conclusions}
\label{sec:summary}

This work presents a systematic study of the spatial and spectral characteristics of the DIG in a sample of 19 nearby ($D < 20$~Mpc) star-forming galaxies spanning a range of ${\sim}1.5$~dex in stellar mass ($9.4 < \log(M_\star/M_\odot) < 11$) at an average spatial resolution of ${\sim}50$~pc. Our data consist of optical IFS obtained in the context of the PHANGS--MUSE Large Programme. Unlike large IFS surveys of nearby galaxies (e.g. CALIFA, SAMI, and MaNGA), which have a resolution of approximately a kiloparsec, we were able to resolve the average distance between \hii\ regions (${\sim}100{-}200$~pc) and therefore spatially isolate \hii\ regions from the surrounding DIG. 

We defined \hii\ regions from \ha\ line emission maps by using the \textsc{HIIphot} algorithm \citep{Thilker2000}, complemented with a point-source finder. We defined the DIG as the ionised gas component found outside the \hii\ region masks. In order to study the line ratios of this low-surface-brightness component, we performed a spatial binning based on the observed \SHa, which allowed us to recover fluxes for \nii, \sii, \oiii, and \hb\ above the $3\sigma$ level over almost the entirety of the observed area. We summarise the main results obtained from these data below.

\begin{itemize}
    \item{The distributions of \SHa\ and EW(\ha) in \hii\ regions and in the DIG for the full sample are nearly non-overlapping to within $1\sigma$, with $\log(\SHa / \mathrm{erg~s^{-1}~kpc^{-2}}) = 38.8$ and $\mathrm{EW}(\ha) \sim 16$~\AA\ representing the best boundaries between the two. Individual galaxies, however, show a wide variety in their distributions, and in several cases the tail of the DIG distribution extends to higher \SHa\ and EW(\ha).}
    \item{Six of our 19 galaxies show extended LI(N)ER emission in their central regions, generally associated with $\mathrm{EW}(\ha) < 3$~\AA. These regions are consistent with being predominantly ionised by HOLMES. On integrated scales, however, the fraction of the total \ha\ emission due to HOLMES is small within our sample (2\% on average, going up to 9\% in the most extreme galaxy, NGC~1433).}
    \item{We model the propagation of leaking radiation from \hii\ regions with a simple thin-slab model. The model takes geometric dilution of the radiation field into account and uses an effective absorption coefficient ($k_0$) to parametrise the mean free path ($\lambda_0 = 1/k_0$) of the ionising photons. Fitting this model to the DIG \ha\ emission in our sample, we obtain an average best-fit mean free path of $1.9$~kpc. Despite relatively large $\chi^2_\mathrm{dof}$, we find that in all galaxies the model has a well-defined $\chi^2_\mathrm{dof}$ minimum as a function of $k_0$ and provides a good qualitative representation of the data at all \SHa\ levels. Our approach demonstrates that the observed \ha\ emission in the DIG is consistent with being powered by leaking radiation, after subtracting the small contribution to the \ha\ flux due to HOLMES.}
    \item{If we assume that the DIG disc has a thickness comparable to the ionised gas scale heights measured from the literature (1~kpc), then we can estimate the escape fraction from \hii\ regions in the context of the leakage model we have developed. We find a median value of the escape fraction of 40\% for our sample. This value is in good agreement with the more naive estimate of the escape fraction obtained by directly comparing the \ha\ flux inside the \hii\ region masks with that outside of them, which results in a median escape fraction of 37\%. }
    \item{In the leaking radiation model, the ionisation parameter is predicted to decrease following the dilution of the ionising photon flux. Assuming a roughly constant electron density, this is reflected in the decrease in \SHa. Metallicity gradients complicate the analysis of line ratios as a function of \SHa. In order to minimise the effect of metallicity, we consider changes in line ratios within radial bins, on average 1.5~kpc in width. In accordance with previous work, we find that low-ionisation line ratios (\sii/\ha, \nii/\ha, and \oi/\ha) increase with decreasing \SHa. Photoionisation models with a 2~Myr SSP as input are able to reproduce these trends and the observed values of the line ratios assuming that decreasing \SHa\ is linked to a decrease in $\log{U_0}$. Different radial bins do not show large differences in \sii/\ha\ and \oi/\ha, because the effect of metallicity on these line ratios is relatively small. \nii/\ha, on the other hand, is found to be systematically higher in the inner regions with respect to the outer regions at fixed \SHa, demonstrating the role of metallicity in setting the value of this line ratio in the DIG.}
    \item{Models of leaking radiation predict that an increase in the ionisation parameter leads to an increase in \oiii/\hb, in clear tension with the observation of the flat or decreasing trend of \oiii/\hb\ with increasing \SHa. A second class of ionising sources with a harder spectrum is needed in order to explain the high observed value of \oiii/\hb\ in the DIG. Moreover, the relative importance of this second component must increase with distance from \hii\ regions in order to reproduce the observed trends with \SHa. We demonstrate that the predicted ionising flux and spectral hardness of HOLMES make them the natural candidate to explain the observations. We find that the line ratios in the DIG can be explained considering a mixing sequence between HOLMES and leaking radiation.} 
    \item{For a fractional contribution of HOLMES to the total ionising photon budget larger than about 10\%, our models extend above the typical demarcation lines used to define the star-forming sequence in the BPT diagram. The mixing sequence spans the full range of line ratios observed in the DIG in our sample and also explains the trend between the position in the BPT diagram and the EW(\ha).}
    \item{Central regions from five of the more massive galaxies in our sample deviate from the average line ratio versus \SHa\ relations, showing systematically higher line ratios at fixed \SHa. These regions correspond to areas of low EW(\ha) and LI(N)ER-like emission in the BPT diagram. These observed properties are naturally explained by a locally dominant contribution of HOLMES to the ionisation budget.}
\end{itemize}

Based on the understanding of the DIG as a mixing sequence between leaking radiation from \hii\ regions and HOLMES, we aim to develop a detailed photoionisation modelling framework for the DIG of star-forming galaxies. Such models, which need to be tested against high-spatial-resolution data, would be a powerful tool for interpreting emission-line ratios observed on kiloparsec scales at both low and high redshift.

\begin{acknowledgements}

This work has been carried out as part of the PHANGS collaboration.

This work is based on observations collected at the European Organisation for Astronomical Research in the Southern Hemisphere under ESO programme IDs 1100.B-0651, 095.C-0473, and 094.C-0623.
FB acknowledges funding from the ESO-Garching fellowship during the earlier stages of development of the project. FB also thanks ASTRO3D for supporting a research stay at ICRAR, which was fundamental to the development of this project.
We thanks the referee for their insightful suggestions, which improved the quality of this work.
ES, FS, HAP, and TGW acknowledge funding from the European Research Council (ERC) under the European Union’s Horizon 2020 research and innovation programme (grant agreement No. 694343).
RSK and SCOG acknowledge financial support from the German Research Foundation (DFG) via the Collaborative Research Center (SFB 881, Project-ID 138713538) `The Milky Way System' (subprojects A1, B1, B2, and B8). They also acknowledge funding from the Heidelberg Cluster of Excellence STRUCTURES in the framework of Germany's Excellence Strategy (grant EXC-2181/1 - 390900948) and from the European Research Council via the ERC Synergy Grant ECOGAL (grant 855130). 
EC acknowledges support from ANID project Basal AFB-170002. 
KK gratefully acknowledges funding from the German Research Foundation (DFG) in the form of an Emmy Noether Research Group (grant number KR4598/2-1, PI Kreckel). 
MB gratefully acknowledges support by the ANID BASAL project FB210003.
ATB would like to acknowledge funding from the European Research Council (ERC) under the European Union's Horizon 2020 research and innovation programme (grant agreement No.726384/Empire).
MC and JMDK gratefully acknowledge funding from the Deutsche Forschungsgemeinschaft (DFG, German Research Foundation) through an Emmy Noether Research Group (grant number KR4801/1-1) and the DFG Sachbeihilfe (grant number KR4801/2-1), as well as from the European Research Council (ERC) under the European Union's Horizon 2020 research and innovation programme via the ERC Starting Grant MUSTANG (grant agreement number 714907).
The work of AKL was partially supported by the National Science Foundation (NSF) under Grants No.1615105, and 1653300. 
Science-level MUSE mosaicked datacubes and high-level analysis products (e.g. emission line fluxes) are provided via the ESO archive phase 3 interface\footnote{\url{https://archive.eso.org/scienceportal/home?data_collection=PHANGS}}. A full description of the the first PHANGS data release is presented in \cite{Emsellem2021}. The \hii\ region catalogue used in this work will be made available in a forthcoming publication. 

\end{acknowledgements}

%
\bibliographystyle{aa} 
\bibliography{library4} 

\begin{appendix}

\section{Atlas of \texorpdfstring{\ha}{Halpha} and \texorpdfstring{\hii}{HII} region mask images for the full sample}
\label{app:atlas}

\begin{figure*}
        \centering
        \includegraphics[width=0.90\textwidth, trim=0 50 0 0, clip]{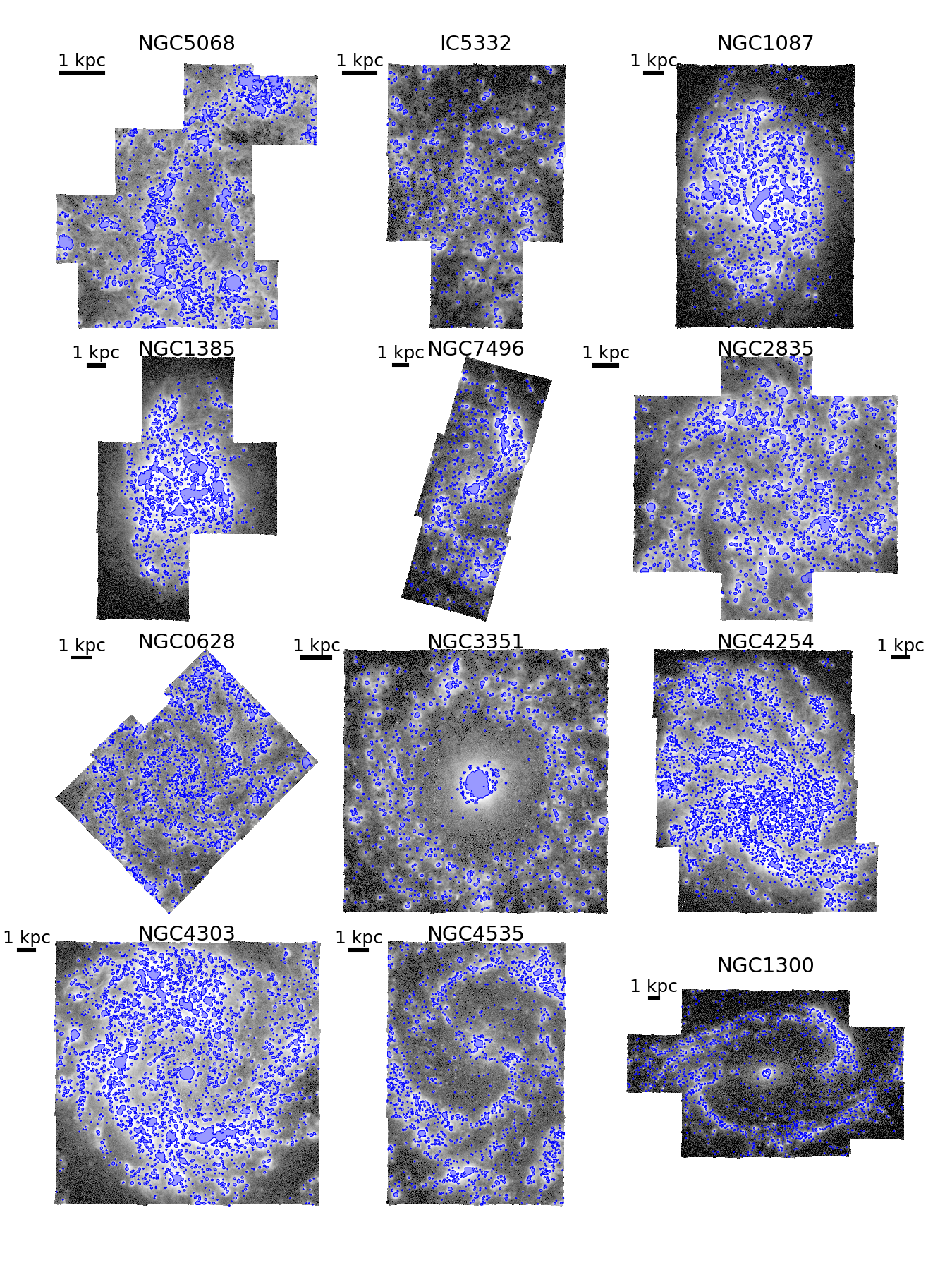}
        \caption{{Ionised nebula masks (blue) superimposed on the $\bm{\SHa}$ maps (grey scale).} Galaxies are ordered by increasing stellar mass from top left to bottom right.}
        \label{HII_region_masks}
\end{figure*}

\begin{figure*}
        \centering
        \includegraphics[width=0.90\textwidth, trim=0 50 0 0, clip]{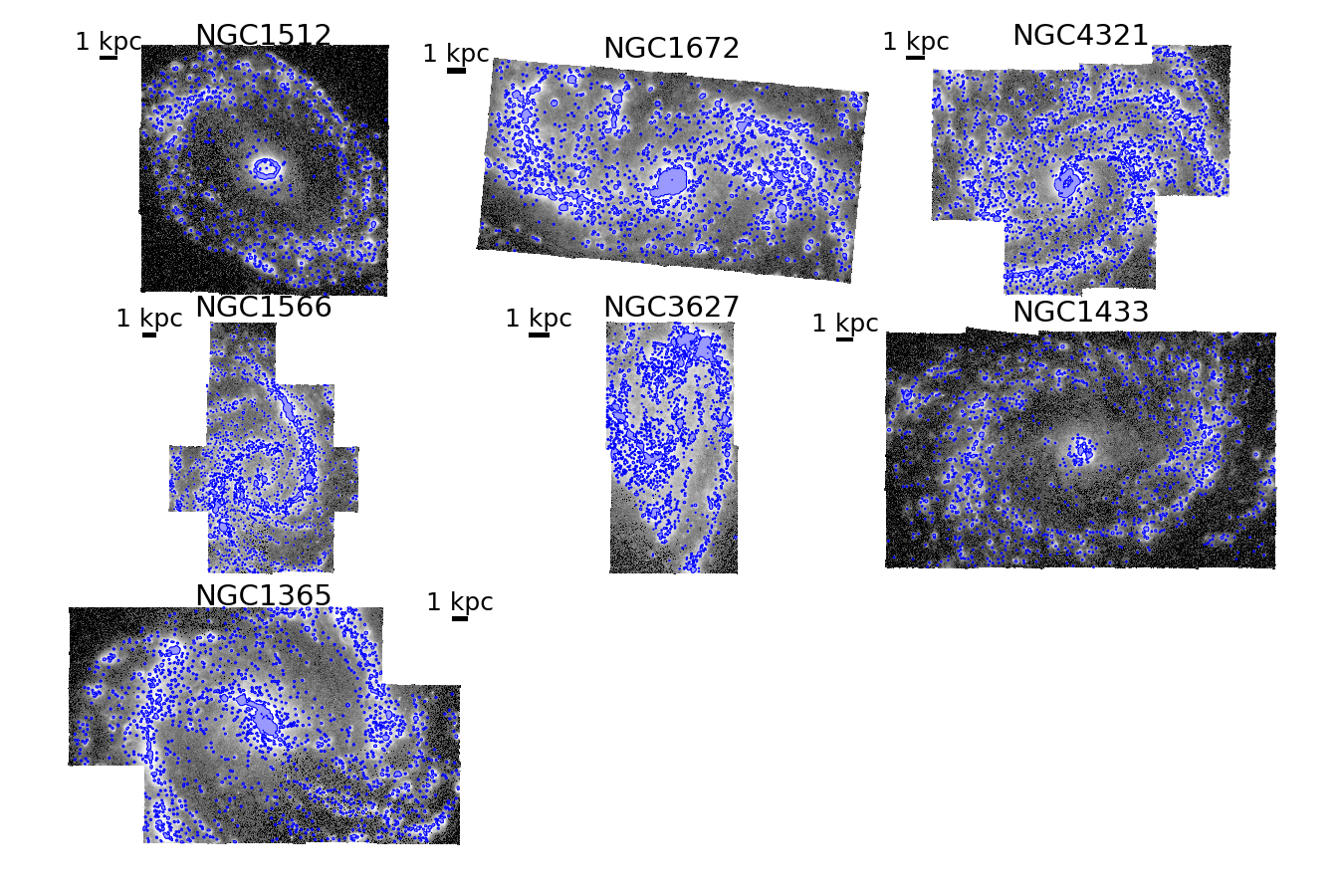}
        \caption{{Same as Fig.~\ref{HII_region_masks}.}}
        \label{HII_region_masks2}
\end{figure*}

In this Appendix, we show \ha\ maps for all PHANGS--MUSE galaxies at their native (spaxel) resolution, with the \hii\ region masks superimposed in blue (Fig. \ref{HII_region_masks}-\ref{HII_region_masks2})

\section{Derivation of the thin-slab model solution for the propagation of ionising radiation in the DIG}
\label{app:spatial_model}

One may model the propagation of ionising radiation in a thin disc filled with a  population of opaque clouds with an effective attenuation law, described by the usual radiative transfer equation,
\begin{equation}
    \frac{d I}{d r} = - I \, k_{0}~,
\end{equation}
where $I(r)$ is the ionising photon -- Lyman continuum (LyC) -- surface brightness at distance $r$ from the ionising source, and $k_{0}$ an effective absorption coefficient, which we take as a free parameter. The ionising luminosity at position $r$ is then given by $I(r) = I_0 \, \mathrm{e}^{-\tau(r)}$, where $I_0 = I (r=0) = F^{\hii} / 4\pi $ and $F^{\hii}$ is the \hii\ region ionising photon flux and $\tau(r) = \int k_{0} \, dr$ is the optical depth. 

Considering a pixel with linear size $\Delta r$, the amount of ionising radiation absorbed within it, which will be proportional to the \ha\ recombination line luminosity, is given by 
\begin{equation}
   \Delta I = I(r)  - I(r+\Delta r ) = I_0 \, \mathrm{e}^{-\tau(r)} \, (1 - \mathrm{e}^{-k_{0} \Delta r}).
   \label{eq:abs}
\end{equation}
In a thin slab of height $\Delta h$, the flux $F(r)$ absorbed in a pixel of size $\Delta  r$ is given by
\begin{equation}
   F(r) = \Delta I \, \Delta \Omega = \Delta I \frac{\Delta r \, \Delta h}{r^2} = F^{\rm \hii} \frac{\Delta r \, \Delta h \, \mathrm{e}^{-\tau(r)}}{4 \pi r^2} (1- \mathrm{e}^{-k_{0} \Delta r}).
\end{equation}
Summing over all the sources of ionising photons (\hii\ regions) and dividing by the pixel area $(\Delta r)^2$, one arrives at Eq.~\ref{eq:model}, noting that conversion factors between ionising photon and \ha\ flux are applied to both sides of Eq.~\ref{eq:model}, and therefore cancel out.

\section{Photoionisation models: Young and old star models with Cloudy v17.02 and FSPS spectra with MIST isochrones}
\label{app:photomodels}

In this paper we have presented photoionisation models based on the \textsc{fsps} code \citep{Conroy2009} and MIST isochrones \citep{Choi2016, Dotter2016}, using \textsc{cloudy} v17.02 \citep{Ferland2017}. These models update and expand the works of \cite{Byler2017,Byler2018, Byler2019}, who used \textsc{cloudy} v13.03. \textsc{cloudy} v>17.00 include a new treatment of the $\rm S^{2+} \rightarrow S^{+}$ dielectronic recombination coefficient. In short, dielectronic recombination ($\rm S^{2+} + e^+ \rightarrow S^{+**} \rightarrow S^{+} + h\nu$) is the main recombination pathway for doubly ionised sulphur under nebular conditions. Unfortunately the necessary atomic data to accurately calculate the dielectronic recombination rate for $\rm S^{2+}$, or other third row elements are not available. \textsc{cloudy} v<17.00 therefore used the charge-normalised mean dielectronic recombination rates for the C, N and O as a guess for the dielectronic recombination for sulphur \citep{Ali1991}. From v17.0 onwards, a new empirically derived formalism for the dielectronic recombination of $\rm S^{2+}$ is included in \textsc{cloudy} \citep{Badnell2015}. The net effect of the change is an increase in the predicted flux of the \sii\ lines, and, in particular, an increase in the \sii/\siii\ ratio of ${\sim}20{-}50\%$ at fixed \oii/\oiii\ (see \citealt{Badnell2015}, their Fig.~8). We considered this update significant enough in the context of this work, and we therefore proceeded to compute models based on almost identical input physics to \cite{Byler2018} but using \textsc{cloudy} v17.02.

In particular, we generate input spectra for a 2~Myr and a 10~Gyr SSP using \textsc{fsps} v3.1 and its python wrappings, \textsc{python-fsps}\footnote{\url{https://github.com/dfm/python-fsps}}. The 2~Myr SSP is chosen because the slope of the ionising UV spectrum at this age is roughly consistent with that of a continuous star-formation model once it reaches equilibrium \citep{Byler2017}. The input spectra are converted into the format required for input into \textsc{cloudy} using the ad hoc script provided in \textsc{cloudy-fsps}\footnote{\url{https://github.com/nell-byler/cloudyfsps}}. The SSPs are generated using the default settings in \textsc{fsps}, MIST isochrones and a \cite{Kroupa2001} initial mass function. \textsc{fsps} makes use of O and B~star spectra generated with WMBasic \citep{Pauldrach2001}, Wolf--Rayet stars are taken from the spectral library of \cite{Smith2002}, and post-AGB stars are modelled with non-LTE model spectra from \cite{Rauch2003}. The MILES stellar library is used for other stellar types. We note that, while the choice of the MILES stellar library sets us apart from the choices made in the Byler papers (BaSEL was used in \citealt{Byler2017} and a new theoretical library still in preparation in \citealt{Byler2018, Byler2019}), we do not expect this to have a significant effect on the predicted line fluxes, which are determined by the choice of library for the hot stars.

The gas-phase abundance model is identical to \cite{Byler2018}. The solar abundance model of \cite{Asplund2009} is used together with the depletion factors by \cite{Dopita2013}. We scale the abundances of carbon and nitrogen with metallicity according to the fitting function of \cite{Byler2018} (their Eqs.~1 and~2; see also the discussion in their Appendix~B), which are based on existing measurements of these abundance ratios in \hii\ regions. The other heavy elements are scaled linearly with metallicity. Unlike the fiducial Byler models, we include ISM grains in our \textsc{cloudy} runs. They  are found to have a small impact on line ratios, but with the largest impact on the high-metallicity high-$\log{U_0}$ models (see discussion in \citealt{Byler2017}, their Sect.~6.3).

In the case of young stars, the ionising  spectrum has a significant dependence on the metallicity of the input spectrum, since lower-metallicity stars have harder spectra. Our 2~Myr models are computed assuming the same metallicity for stars and gas. For the 10~Gyr model there is no physical reason to assume the gas clouds to have the same metallicity as the stars. \cite{Byler2019} find that, however, post-AGB ionising spectra have nearly identical shapes at all metallicities. We therefore arbitrarily fix the metallicity of the input SSP to solar for all the 10~Gyr models.

Models were run in each case for six values of gas-phase metallicity (${\rm [Z/H]} = [-0.6, -0.4, -0.2, 0., 0.2, 0.4]$) and eight values of ionisation parameter ($\log U  = [-5, -4.5, -4, -3.5,-3,-2.5,-2,-1]$) using \textsc{pyCloudy}. The \textsc{cloudy} computation was stopped when the temperature drops below $1000$~K or the ionised fraction below $0.01$. We verified that model runs with \textsc{cloudy} 13.03, no grains, and the old abundance model described in \cite{Byler2017} reproduce their model predictions based on the MIST isochrones well.

Both the 2~Myr and the 10~Gyr model grids are made publicly available\footnote{\url{https://github.com/francbelf/python_izi/tree/master/grids}}.

\end{appendix}

\end{document}